\documentclass[12pt]{article}

\usepackage[dvips]{graphicx}

\newcommand{\nn}{\nonumber}
\newcommand{\be}{\begin{equation}}
\newcommand{\ee}{\end{equation}}
\newcommand{\ba}{\begin{eqnarray}}
\newcommand{\ea}{\end{eqnarray}}
\newcommand{\ci}[1]{\cite{#1}}
\def\={\,=\,}

\def\vd{\Delta_\perp}
\def\vo{{\bf 0}_\perp}

\newcommand{\LQCD}{\Lambda_{\rm{QCD}}}
\def\als{\alpha_s}
\def\ale{\alpha_{\rm em}}

\def\gev{\,{\rm GeV}}

\def\taub{\bar{\tau}}
\newcommand{\tw}{\textwidth}                          
\newcommand{\req}[1]{(\ref{#1})}

\def\eps{\epsilon}
\def\veps{\varepsilon}

\begin{document}
\thispagestyle{empty}
\begin{flushright}
RBI-ThPhys-2022-46\\
WUB/22\\
November, 17 2022\\[20mm]
\end{flushright}

\begin{center}
  {\Large\bf Transition GPDs and exclusive\\[0.3em]
    electroproduction of $\pi-\Delta(1232)$ final states}\\[0.3em]
\vskip 15mm
P.\ Kroll \\[1em]
{\small {\it Fachbereich Physik, Universit\"at Wuppertal, D-42097 Wuppertal,
Germany}}\\[1em]
K.\  Passek-Kumeri\v{c}ki  \\[1em]
{\small {\it Division of Theoretical Physics, Rudjer Bo\v{s}kovi\'{c} Institute, 
HR-10002 Zagreb, Croatia}}\\

\end{center}

\begin{abstract}
  We investigate exclusive electroproduction of $\pi^-\Delta^{++}$ within the handbag approach
  in which the helicity amplitudes factorize into generalized parton distributions (GPDs)
  and hard partonic subprocesses. We define the $p-\Delta$ transversity GPDs while
  the helicity non-flip GPDs are taken from the literature. For the numerical estimates of
  observables we utilize large-$N_C$ results in order to relate a
  few of the $p-\Delta$ GPDs to the proton-proton ones and neglect all other GPDs. In the
  calculation of the twist-2 and twist-3 subprocess amplitudes we take into account quark
  transverse momenta in combination with Sudakov suppressions. The partial cross sections
  for $\gamma^*p\to \pi^-\Delta^{++}$ are predicted in the large-$N_C$ limit.
\end{abstract}  
%%%%%%%%%%%%%%%%%%%%%%%%%%%%%%%%%%%%%%%%%%%%%%%%%%%%%%%%%%%%%%%%%%%%%%%%%%%%%%%%%%%%
\section{Introduction}
%%%%%%%%%%%%%%%%%%%%%%%%%%%%%%%%%%%%%%%%%%%%%%%%%%%%%%%%%%%%%%%%%%%%%%%%%%%%%%%%%%%%
In the last 25 years hard exclusive reactions have found much interest.
The theoretical description of such processes, the handbag approach, is based on 
factorization of the process amplitudes into hard partonic subprocesses and soft hadronic
matrix elements, parameterized as GPDs. This factorization property has been shown 
to hold rigorously to leading-twist accuracy in the generalized Bjorken regime  of 
large photon virtuality, $Q^2$, and large invariant mass of the hadrons in the final 
state, $W$, but fixed Bjorken-x, $x_B$, and small Mandelstam-$t$ \ci{rad96,ji96,collins97}.
However, there is no theoretical estimate of the strength of subasymptotic power corrections.
In so far, the validity of the asymptotic leading-twist result in a given range of $Q^2$
and $W$ is to be regarded as an assumption. The strength of power corrections have to be
extracted from the analysis of relevant data. Still the handbag approach with occasional
power corrections, has been successfully applied to electroproduction of vector (e.g.\
$\rho^0, \phi, \omega$) and pseudoscalar (e.g.\ $\pi, \eta, \eta'$) mesons, see the reviews
\ci{belitsky,diehl03} and references therein. These processes require the diagonal
proton-proton GPDs. Octet-octet transition GPDs occur for instance in kaon electroproduction.
SU(3) flavor symmetry relates these GPDs to the $p-p$ ones provided that symmetry breaking
effects are ignored \ci{frankfurt}. This, for instance, allows an analysis of the
sparse kaon data (see \ci{carmignotto} and references therein) within the handbag approach
\ci{kroll19}. The situation for the octet-decuplet GPDs is more complicated. Only in the
large-$N_C$ limit these GPDs - or at least some of them - can be related to the $p-p$ ones
\ci{belitsky,frankfurt99}.

Since in the near future data on the exclusive $\pi\Delta$ channels will come from the Jefferson
Lab we think it timely to analyze such processes. First, still preliminary, data, namely the
beam spin asymmetry for $\pi^-\Delta^{++}(1232)$, have already been shown on conferences
\ci{diehl22}. Here in this article we are going to analyze the process
$\gamma^*p\to \pi^-\Delta^{++}(1232)$ in full analogy to $\gamma^*N\to \pi N'$ \ci{GK5,GK6}. It
however turned out for the latter process that the naive asymptotic leading-twist result 
is not readily applicable in the range of kinematics accessible to these experiments.
In fact, large power corrections are required to the asymptotically dominant 
amplitudes for longitudinally polarized photons. There are also strong contributions from
transversally polarized photons although they are asymptotically suppressed by $1/Q^2$ 
in the cross sections. In some cases, as for instance for $\pi^0$ electroproduction 
\ci{defurne}, the contributions from transversely polarized photons are even dominant. 
In \ci{GK5,GK6}, see also the review \ci{kroll14}, a generalization of the handbag approach
has been developed which allows to model such power corrections. The decisive point is to retain
the  quark transverse momenta in the subprocess and to apply Sudakov suppressions. Implicitly,
this way the transverse size of the meson is taken into account~\footnote{
     The role of quark transverse momenta and the transverse size of the meson has also been
     discussed in \ci{frankfurt95} in the case of light vector-meson production at HERA kinematics.}.
The contribution from transversely polarized photons are modeled by a combination of
transversity GPDs and a twist-3 pion wave function. This twist-3 contribution is large
in the case of pions since it is proportional to a mass parameter, $\mu_\pi$, which is not the usual
mass of the pion. It is rather enhanced by the chiral condensate
\be
\mu_\pi\=\frac{m_\pi^2}{m_u+m_d}
\label{eq:mass-parameter}
\ee
by means of the divergence of the axial vector current~\footnote{Twist-3 effects may also be
  generated by twist-3 GPDs in combination with a twist-2 meson wave function \ci{schafer01}.
  However, for these GPDs there is no similar enhancement known. Therefore, such contributions
  are expected to be small and neglected here.}.
The masses $m_u$ and $m_d$ are the current-quark masses of the pion's valence quarks. The mass
parameter is large, about $2\,\gev$ at the initial scale $\mu_0=2\,\gev$. As in \ci{GK5,GK6,GK8}
the contribution from the pion pole is treated as a one-bose-exchange term which is much larger
than its leading-twist approximation.

The plan of the paper is the following: After some kinematical preliminaries (Sect.\ 2) we calculate
the matrix elements, $A^\Delta$, for the helicity non-flip $p-\Delta$ GPDs and the
$\gamma^*p\to \pi^-\Delta^{++}$ helicity amplitudes (Sect.\ 3). In the next section we define
the transversity GPDs and calculate likewise the $A^\Delta$ and the helicity amplitudes for them. Sect.\ 5
is devoted to a study of the contribution from the pion pole and, in Sect.\ 6, we discuss the GPDs
in the large-$N_C$ limit. In Sect.\ 7 the parameterization of the GPDs and the calculation of the
subprocess amplitudes is presented. Our predictions for the $\gamma^*p\to \pi^-\Delta^{++}$ partial
cross sections are shown and discussed in Sect.\ 8. Finally, in Sect.\ 9, we give a summary.
%%%%%%%%%%%%%%%%%%%%%%%%%%%%%%%%%%%%%%%%%%%%%%%%%%%%%%%%%%%%%%%%%%%%%%%%%%%%%%%%%%%%%%%%%%%
%\section{Handbag factorization}
%%%%%%%%%%%%%%%%%%%%%%%%%%%%%%%%%%%%%%%%%%%%%%%%%%%%%%%%%%%%%%%%%%%%%%%%%%%%%%%%%%%%%%%%%%%%%%%%%
\section{Kinematical preliminaries}
\label{sec:kinematics}
%%%%%%%%%%%%%%%%%%%%%%%%%%%%%%%%%%%%%%%%%%%%%%%%%%%%%%%%%%%%%%%%%%%%%%%%%%%%%%%%%%%%%%%%%%%%
We are interested in the hard exclusive process
\be
\gamma^*(q,\mu) p (p,\nu)\to \pi^-(q',\mu') \Delta^{++} (p',\nu')
\ee
where the labels in the brackets denote the momenta of the particles and their light-cone helicities.
In light-cone components the momenta are defined as
\ba
p&=& \left[(1+\xi) P^+,\,\frac{m^2+\vd^2/4}{2(1+\xi) P^+}\,, -\frac{\vd}{2}, 0 \right]\nn\\
p'&=& \left[(1-\xi) P^+,\,\frac{M^2+\vd^2/4}{2(1-\xi) P^+}\,,\phantom{-} \frac{\vd}{2}, 0
           \right]\nn\\    
  q&=& \left[\eta(1+\xi) P^+,\frac{-Q^2+\vd^2/4}{2\eta(1+\xi) P^+}, \frac{\vd}{2},0 \right]
\label{eq:ji-momenta}
\ea
where $m$ and $M$ denote the mass of the proton and the $\Delta(1232)$, respectively.
The mass of the pion is neglected except in the pion pole term, see below. The negative
of $\eta$ equals the Bjorken variable, $x_B$, up to corrections of order $m^2/Q^2$ and $\vd^2/Q^2$.
It is convenient to introduce an average baryon momentum, $P$, and a momentum transfer, $\Delta$:
\be
P\=\frac12\,\big( p+p'\big)\,, \qquad \Delta\=p'-p\,.
\ee
The skewness is defined by the ratio of light-cone plus components of $\Delta$ and $P$
\be
\xi\=- \frac{\Delta^+}{2P^+}
\ee
and is related to Bjorken-$x$ by
\be
\xi\=\frac{x_B}{2 - x_B}\,.
\ee
This relation strictly holds for $Q^2\to \infty$ but we neglect here the corrections of order $1/Q^2$.
Mandelstam $t$ is given by
\be
t\= \Delta^2\= t_0 - \frac{\vd^2}{1-\xi^2}
\ee
where $t_0$ is the minimal value of $t$ implied by the positivity of $\vd^2$
\be
t_0\=- \frac{2\xi}{1-\xi^2}\,\Big[(1+\xi)(M^2-m^2) + 2\xi m^2\Big]\,.
%t_0\=- \frac{2\xi}{1-\xi^2}\,\Big[(1+\xi)M^2-(1-\xi)m^2\Big]\,.
\ee
For convenience we will frequently use in the following a variable $t'$
defined by
\be
t'= t - t_0 = - \Delta_\perp^2/(1-\xi^2)\,.
\ee
For later use we also define the two quantities
\be
\kappa_\pm \= (1+\xi) M \pm (1-\xi) m\,.
\ee

%%%%%%%%%%%%%%%%%%%%%%%%%%%%%%%%%%%%%%%%%%%%%%%%%%%%%%%%%%%%%%%%%%%%%%%%%%%%%%%%%%%%%%%%%%%%
\section{The helicity non-flip $p-\Delta$ transition GPDs}
%%%%%%%%%%%%%%%%%%%%%%%%%%%%%%%%%%%%%%%%%%%%%%%%%%%%%%%%%%%%%%%%%%%%%%%%%%%%%%%%%%%%%%%%%%%
We will make use of the helicity non-flip $p-\Delta$ GPDs defined by Belitsky and Radyushkin \ci{belitsky}.
There are four even-parity transition GPDs~\footnote{
  In \ci{belitsky} $\Delta$ is defined as $p-p'$ and $P$ as $p+p'$.}
\ba
  \lefteqn{
    \frac12 \int\frac{dz^-}{2\pi} e^{ix P^+z^-}\, \langle \Delta^{++}(p',\nu')|\bar{u}(-z/2)\gamma^+ d(z/2)
              |p(p,\nu)\rangle\left|_{z^+=0,z_\perp=0} =\right. }  \nn\\
     && \frac1{2P^+}\,\bar{u}_\delta(p',\nu')\left\{ \frac{\Delta^\delta n^\mu - \Delta^\mu n^\delta}{m} 
              \Big(\gamma_\mu G_1(x,\xi,t)   \right.\nn\\
&& \left.  + \frac{P_\mu}{m} G_2(x,\xi,t)
                   + \frac{\Delta_\mu}{m} G_3(x,\xi,t)\Big)  
         + \frac{\Delta^+\Delta^\delta}{m^2} G_4(x,\xi,t)\right\} \gamma_5 u(p,\nu)  
   \label{eq:even-pD-GPDs}
  \ea
  and four odd-parity ones
  \ba
  \lefteqn{     
   \frac12  \int\frac{dz^-}{2\pi} e^{ix P^+z^-}\,\langle \Delta^{++}(p',\nu')|\bar{u}(-z/2)\gamma^+\gamma^5 d(z/2)
     |p(p,\nu)\rangle\left|_{z^+=0,z_\perp=0} = \right. } \nn\\
    && \frac1{2P^+}\,\bar{u}_\delta(p',\nu')\left\{ \frac{\Delta^\delta n^\mu - \Delta^\mu n^\delta}{m} 
           \Big(\gamma_\mu \tilde{G}_1(x,\xi,t)     \right.\nn\\
   &&\left. + \frac{P_\mu}{m} \tilde{G}_2(x,\xi,t)\Big)
                   + n^\delta \tilde{G}_3(x,\xi,t)  
                   + \frac{\Delta^+\Delta^\delta}{m^2} \tilde{G}_4(x,\xi,t)\right\} u(p,\nu)
 \label{eq:odd-pD-GPDs}
  \ea
  where the vector $n$ is
  \be
  n\= \Big[0,1,\vo\Big]\,.
  \ee
  In comparison with the diagonal proton-proton GPDs, see for example \ci{diehl01}, an extra $\gamma_5$
  is introduced in order to match the parity of the $\Delta(1232)$. The $\Delta(1232)$ is considered as a
  stable particle. Therefore, one can apply time-reversal invariance to show that the $p-\Delta$ transition
  GPDs are real-valued functions \ci{diehl03}. In the definitions \req{eq:even-pD-GPDs} and \req{eq:odd-pD-GPDs}
  $u(-z/2)$ and $d(z/2)$ denote quark field operators of a specified  flavor and we work in the light-cone gauge $A^+=0$.
  The GPDs also depend on the
  factorization scale which, for convenience, is suppressed in \req{eq:even-pD-GPDs}, \req{eq:odd-pD-GPDs}
  and in the following. The $u_\delta(p',\nu')$ in the above relations denotes a Rarita-Schwinger spinor
  for the $\Delta(1232)$. It satisfies the Dirac equation and is subject to the subsidiary conditions
  \be
  p'_\mu u^\mu(p',\nu')\=0\,, \qquad \gamma_\mu u^\mu(p',\nu')\=0\,.
  \label{eq:conditions}
  \ee
  The Rarita-Schwinger spinors can, for instance, be found in \ci{compilation}.
  The lowest moments of the GPDs $G_1, G_2$ and $G_3$ are related to the proton-$\Delta$ transition form factors,
  see \ci{belitsky}.

  Isospin symmetry relates the $p-\Delta^{++}$ GPDs to those of other $\Delta(1232)$ states \ci{belitsky}
  \be
  G^{ud}_{p\Delta^{++}} \= - \frac{\sqrt{3}}{2}\,G^{uu-dd}_{p\Delta^+} \= -\sqrt{3}\,G^{du}_{p\Delta^0}\,.
\label{eq:isospin}
\ee
As an special exception we have indicated the flavor content of the GPDs in this relation. Thus, for
instance, $G^{ud}_{p\Delta^{++}}$ is a GPD for which an $d$-quark is emitted from the proton and $u$-one
reabsorbed forming the $\Delta^{++}$. The relations \req{eq:isospin} hold for all eight helicity
non-flip GPDs as well as for the transversity GPDs which will be defined  below. SU(3) flavor symmetry
also relates the $p-\Delta$ GPDs  to other octet-decuplet transition GPDs.

\begin{figure}
      \centering
  \includegraphics[width=0.4\tw]{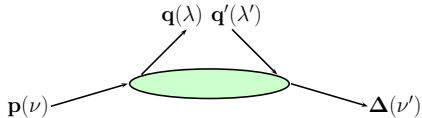}
      \caption{The matrix element representing a proton-$\Delta$ GPD. }
      \label{fig:matrix-element}
\end{figure}

We are interested in matrix elements for emitted and reabsorbed partons with definite light-cone helicity.
According to Diehl \ci{diehl01} this can be achieved by considering the following combinations of quark
field operators
  \be
     {\cal O}_{\pm\pm} \= \frac14 \bar{u}(-z/2) \gamma^+ (1 \pm \gamma_5) d(z/2)
  \label{eq:Opp}
  \ee
     where, in the region of $\xi<x<1$, the corresponding matrix elements (see Fig.\ \ref{fig:matrix-element})
     \be
     A_{\nu'\lambda \nu\lambda}^\Delta \=  \int\frac{dz^-}{2\pi} e^{ix P^+z^-}\,
                                            \langle \Delta^{++}(p'\nu')|{\cal O}_{\lambda\lambda}|
                                                  |p(p,\nu)\rangle\left|_{z^+=0,z_\perp=0} \right. 
     \label{eq:pp}
     \ee
describe the emission and reabsorption of on-shell quarks with helicity $\lambda$.  In other regions of $x$
one has, if necessary, to reinterpret an outgoing (incoming) quark with helicity $\lambda$ as incoming
(outgoing) quark with helicity $-\lambda$.

Using the definitions \req{eq:even-pD-GPDs} and \req{eq:odd-pD-GPDs} one can readily work out the matrix
elements for the kinematics described in Sect. \ref{sec:kinematics}~\footnote{
        These results hold in any frame provided that ${\bf p}$ and ${\bf p}'$ lie in the 1-3 plane %
except for the fact that any $\sqrt{-t'}$ is to be multiplied by ${\rm sign}(P^+\Delta^1-\Delta^+P^1)$ \ci{diehl01}.}:
      \ba
      A^\Delta_{\,3\pm +\pm}&=& +\frac1{2\sqrt{2}} \frac{\sqrt{-t'}}{m^2} \frac1{1-\xi}\,
                  \left\{m(1-\xi^2)(G_1\pm \tilde{G}_1) \right.\nn\\
                 &&\left.\hspace*{0.125\tw}+ \kappa_-\big(\frac12 G_2 - \xi G_3 -\xi G_4\big) \pm
                                           \kappa_+\big(\frac12\tilde{G}_2 -\xi \tilde{G}_4\big)
                                           \right\}\,, \nn\\
   A^\Delta_{-3\pm +\pm}&=& +\frac1{2\sqrt{2}} \frac{t'}{m^2} \sqrt{\frac{1+\xi}{1-\xi}}\,\left\{
   - \frac12 G_2  + \xi G_3 + \xi G_4 \mp \big(\frac12 \tilde{G}_2 - \xi \tilde{G}_4\big)
                                 \right\}\,,\nn\\
    A^\Delta_{\,1\pm +\pm}&=& \frac1{2\sqrt{6}}\,\frac1{m^2M} \frac1{(1-\xi^2)^{3/2}}   \nn\\
    &\times& \left\{(1-\xi^2) m \Big[ (1-\xi^2) t'(G_1\pm\tilde{G}_1)
                            + 2\xi M(\kappa_- G_1\pm \kappa_+\tilde{G}_1)\Big] \right.\nn\\
        &+&\left.\;\frac12 (1+\xi)(1-\xi^2) M t'(G_2\pm \tilde{G}_2)  \right.\nn\\
        &+&\left. \frac12 \Big[\xi(1+\xi)(3-\xi) M^2 + \xi(1-\xi)^2m^2 + (1-\xi^2) t'\Big](\kappa_-G_2 \pm\kappa_+\tilde{G}_2)\right.\nn\\
    &+&\left.\;(1-\xi)(1-\xi^2) m^2 \big(\kappa_-  G_3 \pm \kappa_+\tilde{G}_3\big)
                           + \Big[\xi(1-3\xi)(1+\xi) \kappa_-M^2     \right.\nn\\
    &-&\left. (1-\xi^2)\big(\kappa_-+\xi(1+\xi) M\big)\,t'
             - (1+2\xi)(1-\xi)^2\kappa_- m^2\Big] G_3   \right.\nn\\
    &-&\left.\; \xi(1+\xi)(1-\xi^2) M t'(G_4\pm\tilde{G}_4)  \right.\nn\\
    &&\left.\hspace*{0.2\tw}- \xi( (1-\xi^2) t' + \kappa_-\kappa_+)(\kappa_- G_4
                                                    \pm\kappa_+ \tilde{G}_4)\right\}\,,\nn\\
     A^\Delta_{-1\pm +\pm}&=& \frac1{2\sqrt{6}}\,\frac{\sqrt{-t'}}{m^2 M} \frac1{1-\xi^2}\,\left\{
       -2\xi (1-\xi^2) M m (G_1\pm \tilde{G}_1)  \right. \nn\\
     &+&\left. (1-\xi)(1-\xi^2)m^2 (G_1\mp \tilde{G}_1) \right. \nn\\
       &-&\left. \frac12 (G_2\pm\tilde{G}_2)\Big[(1-\xi^2)t' + m^2\xi(1-\xi)^2 + M^2\xi(1+\xi)(3-\xi)\Big]
                                             \right.\nn\\
      && \left. -\frac12 M(1+\xi)(\kappa_-G_2 \pm\kappa_+\tilde{G}_2) \right. \nn\\
      &-& \left. (1-\xi)(1-\xi^2) m^2(G_3\pm\tilde{G}_3) +   G_3\,\Big[ (1-\xi^2) t' +4 \xi^2 (1+\xi)M^2
                                               \right.\nn\\
      && \left. +(1+2\xi)(1-\xi)^2m^2 - \xi(1-\xi^2) mM\Big] \right.\nn\\
     &+&\left. (G_4\pm\tilde{G}_4)\xi ((1-\xi^2) t'+\kappa_-\kappa_+) \right.\nn\\
     && \left. \hspace*{0.25\tw} + \xi(1+\xi) M (\kappa_- G_4\pm \kappa_+\tilde{G}_4)\,.
     \right\}
     \label{eq:A-long}
     \ea
     Explicit helicities of  photons, nucleons and quarks are labeled by their signs whereas the helicities of
     the $\Delta(1232)$ are denoted by $\pm 3$ and $\pm 1$.

     At leading-twist accuracy only longitudinally polarized photons contribute to the center-of-mass helicity
     amplitudes for the process $\gamma^* p\to \pi^-\Delta^{++}$. The amplitudes read
     \ci{frankfurt99}
     \be
        {\cal M}_{0\nu'0\nu}^{tw2}\= e_0\int_{-1}^1 dx \sum_\lambda {\cal H}^{\pi^-}_{0\lambda 0\lambda}
                                                                 A^{\Delta (3)}_{\nu'\lambda \nu\lambda}\,.
     \ee
     The matrix elements $A^{\Delta (3)}_{\nu'\lambda\nu\lambda}$ appear in the isovector combination, i.e.\ any GPD
     contributes in the flavor combination
     \be
        \tilde{G}_i^{(3)}\= \tilde{G}_i^u-\tilde{G}_i^d\,.
         \label{eq:isovector}
     \ee
     Parity symmetry leads to the following relation for the $\gamma^* q\to \pi^-q$ subprocess amplitudes
     \be
     {\cal H}^\pi_{0-\lambda'-\mu-\lambda} \= - (-1)^{\mu-\lambda + \lambda'} \,{\cal H}^\pi_{0\lambda'\mu\lambda}\,.
     \label{eq:parity-H}
     \ee
     With its help we can write $ {\cal M}_{0\nu'0\nu}^{tw2}$ as
     \be
        {\cal M}_{0\nu'0\nu}^{tw2}\= e_0\int_{-1}^1 dx {\cal H}^\pi_{0+0+}
                               \Big[ A^{\Delta (3)}_{\nu'+ \nu +} - A^{\Delta (3)}_{\nu'- \nu -}\Big]\,.
     \ee
     Inspection of \req{eq:A-long} reveals that only the odd-parity GPDs contribute to pion production. Explicitly,
     the helicity amplitudes read
     \ba
          {\cal M}_{03,0+}^{tw2}&=&\frac{e_0}{\sqrt{2}} \frac{\sqrt{-t'}}{m^2} \frac1{1-\xi} \left\{ 
          (1-\xi^2)\,m\,\langle \tilde{G}_1^{(3)}\rangle 
          + \frac{\kappa_+}{2} \langle \tilde{G}_2^{(3)} \rangle - \xi\kappa_+\langle \tilde{G}_4^{(3)}\rangle
                                    \right\}\,, \nn\\
          {\cal M}_{0-3,0+}^{tw2}&=& -\frac{e_0}{\sqrt{2}} \frac{t'}{m^2} \sqrt{\frac{1+\xi}{1-\xi}} 
          \left\{ \frac12\langle \tilde{G}_2^{(3)} \rangle - \xi\langle \tilde{G}_4^{(3)} \rangle\right\}\,, \nn\\
          {\cal M}_{01,0+}^{tw2}&=& \frac{e_0}{\sqrt{6}} \frac1{m^2M} \frac{1}{(1-\xi^2)^{3/2}}\,
          \left\{(1-\xi^2)m \Big[(1-\xi^2) t' + 2\xi M\kappa_+\Big]\,\langle \tilde{G}_1^{(3)}\rangle \right.\nn\\
          &+&\left.\hspace*{-0.01\tw} \frac12 \Big[(1+\xi)(1-\xi^2)M t' \right.\nn\\
          &+& \left. \Big(\xi(1-\xi)^2m^2+\xi(1+\xi)(3-\xi)M^2 + (1-\xi^2) t'\Big) \kappa_+\Big]\,\langle \tilde{G}_2^{(3)}\rangle\right.\nn\\
          &+& \left. (1-\xi^2)^2(1-\xi) m^2\kappa_+ \langle \tilde{G}_3^{(3)}\rangle \right.\nn\\
          &-&\left. \xi \Big[(1-\xi^2) t'(\kappa_++(1+\xi)M)  + \kappa_-\kappa_+^2\Big]\,
           \langle \tilde{G}_4^{(3)}\rangle \right\}\,,  \nn\\
          {\cal M}_{0-1,0+}^{tw2}&=& \frac{e_0}{\sqrt{6}} \frac{\sqrt{-t'}}{m^2M} \frac1{1-\xi^2}\,
             \left\{- (1-\xi^2)m\Big[2\xi M + (1-\xi)m\Big] \langle \tilde{G}_1^{(3)} \rangle \right.\nn\\
          &-&\left. \frac12 \Big[(1-\xi^2)t'+\xi(1-\xi)^2 m^2 +  (1+\xi)M\kappa_+ \right.\nn\\ 
            &&\left. \qquad  + \xi(1+\xi) (3-\xi) M^2\Big] \,\langle \tilde{G}_2^{(3)} \rangle\right.\nn\\
          &-& \left. (1-\xi)(1-\xi^2) m^2\langle \tilde{G}_3^{(3)} \rangle\right.\nn\\
             &+&\left. \xi\Big[(1-\xi^2)t' + \kappa_-\kappa_+ + (1+\xi)M\kappa_+\Big]\,\langle \tilde{G}_4^{(3)}
                                     \rangle \right\}\,,
           \label{eq:long-amplitudes}
          \ea
          where
          \be
          \langle \tilde{G}_i^{(3)} \rangle \= \int_{-1}^1 dx {\cal H}^{\pi^-}_{0+0+}  \tilde{G}_i^{(3)}\,.
          \label{eq:tw2-convolutions}
          \ee
     
        Parity symmetry for the $\gamma^*p\to \pi^-\Delta^{++}$ helicity amplitudes leads to the relation
        \be
           {\cal M}_{0-\nu'-\mu-\nu}\=(-1)^{\mu-\nu+\nu'}\,{\cal M}_{0\nu'\mu\nu}
           \label{eq:parity-M}
         \ee
        The generalization of these results to mesons in the final state other than the pion is straightforward.

%%%%%%%%%%%%%%%%%%%%%%%%%%%%%%%%%%%%%%%%%%%%%%%%%%%%%%%%%%%%%%%%%%%%%%%%%%%%%%%%%%%%%%%%%%%
\section{The transversity $p-\Delta$ GPDs}
%%%%%%%%%%%%%%%%%%%%%%%%%%%%%%%%%%%%%%%%%%%%%%%%%%%%%%%%%%%%%%%%%%%%%%%%%%%%%%%%%%%%%%%%%%
For the definition of the eight transversity GPDs we have to consider the tensor matrix element
\ba
  \frac12 \int\frac{dz^-}{2\pi} e^{ix P^+z^-}\, \langle \Delta^{++}(p',\nu')|\bar{u}(-z/2)i\sigma^{+j} d(z/2)
    |p(p,\nu)\rangle\left|_{z^+=0,z_\perp=0}    \right.   \nn\\[0.01\tw]
     =   \bar{u}_\delta(p',\nu')\Gamma^{\delta +j}\gamma_5 u(p,\nu)\,.  
\ea
  Here, $\Gamma^{\delta +j}$ is a matrix in the Dirac space being antisymmetric in the Lorentz
  label + and the transverse one $j (=1,2)$. The matrix $\gamma_5$ occurs as a consequence of the spin
  and the parity of the $\Delta(1232)$. At disposal for the construction of $\Gamma$ are the momenta
  $P$, $\Delta$ and $n$ as well as $\gamma^\alpha$ and $\sigma^{\alpha\beta}$ taking into account
  the conditions \req{eq:conditions}. The antisymmetric tensor can also be constructed directly with
  the Rarita-Schwinger spinor. A possible set of transversity GPDs is then
   \ba
   \lefteqn{
     \frac12\,\int \frac{dz^-}{2\pi}\,e^{ixP^+z^-}\,\langle \Delta^{++}(p',\nu')|\bar{u}(-z/2)i\sigma^{+j} 
     d(z/2)|p (p,\nu)\rangle  \left|_{z^+=0,z_T=0} = \right.}\hspace*{0.15\tw} \nn\\
          && \frac{1}{2P^+}\,\bar{u}_\delta(p',\nu')\left[G_{T1} \frac{p^\delta}{m} i\sigma^{+j} 
     + G_{T2}\,p^\delta\, \frac{P^+\Delta^j -\Delta^+ P^j }{m^3} \right.   \nn\\
     &+&\left. G_{T3}\,p^\delta\,\frac{\gamma^+\Delta^j -\Delta^+\gamma^j }{2m^2}
     + G_{T4}\,p^\delta\,\frac{\gamma^+P^j - P^+\gamma^j}{m^2}    \right. \nn\\
     &+&\left.  G_{T5}\, (n^\delta\gamma^j-\gamma^\delta n^j)  + G_{T6}\,\frac{n^\delta \Delta^j-\Delta^\delta n^j}{m} 
                          \right]\,  \gamma_5\, u(p,\nu) \nn\\
   &+&  \frac{1}{2P^+}\,\left[ \quad G_{T7}\, (\bar{u}^+(p',\nu')\gamma^j
                                                - \bar{u}^j(p',\nu')\gamma^+) \right. \nn\\ 
      &+&  \left.   G_{T8}\,  \frac{\bar{u}^+(p',\nu')\Delta^j-\bar{u}^j(p',\nu')\Delta^+}{m}   
                         \right]\,  \gamma_5\, u(p,\nu)
   \label{eq:trans-GPDs}
   \ea
   As the helicity non-flip GPDs the transversity ones are real-valued functions of $x, \xi$ and $t$.
   Any other antisymmetric tensor can be expressed
   as a linear combination of the eight tensors appearing in \req{eq:trans-GPDs}. This can be shown with the
   help of the Dirac equation and the generalized Gordon identities
    \ba
   \bar{u}_\delta(p') i\sigma^{\alpha\beta}(p'\mp p)_\beta u(p)&=&(M\pm m) \bar{u}_\delta(p')\gamma^\alpha u(p)
                                                             - \bar{u}_\delta(p')(p'\pm p)^\alpha u(p)\nn\\
  \bar{u}_\delta(p') i\sigma^{\alpha\beta}(p'\pm p)_\beta\gamma_5 u(p)&=&(M\pm m) \bar{u}_\delta(p')\gamma^\alpha
                                                             \gamma_5 u(p) \nn\\
                                    &&\hspace*{0.2\tw} - \bar{u}_\delta(p')(p'\mp p)^\alpha \gamma_5 u(p)\,.
\label{eq:gordon}                          
\ea
Obviously, the set of GPDs is not unique.

   The matrix elements
   \be
   A_{\nu'-\lambda \nu\lambda}^\Delta\=  \int\frac{dz^-}{2\pi} e^{ix P^+z^-}\,
       \langle \Delta^{++}(p'\nu')|{\cal O}_{-\lambda\lambda}|p(p,\nu)\rangle\left|_{z^+=0,z_\perp=0} \right.
     \label{eq:pm}
   \ee
   where
   \be
      {\cal O}_{\mp\pm}\=\mp\frac{i}{4} \bar{u}(-z/2)\big( \sigma^{+1}\mp i \sigma^{+2}\big) d(z/2)
   \label{eq:Omp}
   \ee
   describe, in the region $\xi<x<1$, the emission of an on-shell quark with helicity $\lambda$ and the
   reabsorption of an on-shell quark with helicity $-\lambda$ \ci{diehl01}. Explicitly these matrix elements read
  \ba
   A^\Delta_{3-,++}&=& -\frac1{4\sqrt{2}} \sqrt{\frac{1+\xi}{1-\xi}}\,\frac{t'}{m^2}\,\left\{ 
   \frac{\kappa_-}{m}\,G_{T2} + (1+\xi)\,( G_{T3} - G_{T4})   \right\}\,, \nn\\
   A^\Delta_{3+,+-}&=&  \frac1{4\sqrt{2}}\, \sqrt{\frac{1+\xi}{1-\xi}}\,\left\{\frac{t'}{m^2}\,\Big[ 
   \frac{\kappa_-}{m}\,G_{T2} + (1-\xi)\,(G_{T3} + G_{T4})\Big]  \right.\nn\\
                 && \left.  + 4(1-\xi)\,G_{T7}  - 4 \frac{\xi}{1+\xi} \frac{\kappa_-}{m}\,G_{T8}\right\}\,, \nn\\
   A^\Delta_{-3-,++}&=&-\frac1{4\sqrt{2}}\,\frac{\sqrt{-t'}}{m}\,\left\{-4(1+\xi)\,G_{T1} + (1+\xi) \frac{t'}{m^2}\,G_{T2}\right.\nn\\
   &+&\left. 2\frac{\kappa_+}{m}\frac{\xi G_{T3} - G_{T4}}{1-\xi}    - 4\xi\,G_{T8}  \right\}\,,\nn\\
   A^\Delta_{-3+,+-}&=& - \frac1{4\sqrt{2}}\,\frac{(-t')^{3/2}}{m^3}\,(1+\xi)\,G_{T2}\,,  \nn\\
   A^\Delta_{1-,++}&=& \frac1{4\sqrt{6}}\,\frac{\sqrt{-t'}}{m}\,\frac1{1-\xi}\,\left\{ - 4 (1-\xi^2)\,G_{T1}
                                                                   \right.\nn\\ 
         &&\left.+ \frac{\kappa_-^2\kappa_++(1-\xi^2) t'(\kappa_-+(1+\xi)M)}{(1+\xi) m^2 M}\,G_{T2} \right. \nn\\
  &+&\left. \frac{\kappa_+\kappa_- +(1-\xi^2)t'}{mM}\,(G_{T3}-G_{T4}) 
                             + 2\frac{\kappa_+}{m}\,(\xi G_{T3}-G_{T4}) \right.\nn\\
  &-&\left.  2(1-\xi)\Big[(1-\xi^2)\frac{m}{M} G_{T5} + (1-\xi) \frac{\kappa_-}{M}\,G_{T6}                                  
                               + \frac{\kappa_-}{M}\,G_{T8} \Big]      \right\}\,, \nn\\
  A^\Delta_{1+,+-}&=& -\frac1{4\sqrt{6}}\,\frac{\sqrt{-t'}}{m}\,\frac1{1+\xi}\,\left\{  
          \frac{\kappa_-^2\kappa_++(1-\xi^2) t'(\kappa_-+M(1+\xi))}{(1-\xi) m^2M}\,G_{T2}   \right. \nn\\
        &+&\left.  \frac{\kappa_-\kappa_+  + (1-\xi^2)t'}{mM}\,\Big[ G_{T3}+G_{T4}\Big]
                    + 2(1-\xi)(1-\xi^2)\frac{m}{M}\,G_{T5}    \right.\nn\\
     &-& \left. 2 (1+\xi)\Big[(1-\xi)\frac{\kappa_-}{M}\,G_{T6}  
        - 2 (1-\xi)\frac{m}{M}\,G_{T7} + \frac{\kappa_-+2\xi M}{M}\,G_{T8}\Big] \right\}\,,\nn\\
  A^\Delta_{-1-,++}&=& \frac1{4\sqrt{6}}\,\frac1{\sqrt{1-\xi^2}}\,\left\{
                      -4\frac{(1-\xi^2)t'+\kappa_-\kappa_+}{mM}\,G_{T1}  \right.\nn\\
         &&\left.  +  \frac{t'}{m^3M}\Big[\kappa_-(\kappa_+ +(1+\xi)M)+(1-\xi^2) t'\Big]\,G_{T2}
                         \right.\nn\\
             &+&\left. \frac{(1+\xi)^2  t'}{m^2}\,(G_{T3}-G_{T4}) 
     + 2\frac{\kappa_+(\kappa_-\kappa_+ + (1-\xi^2) t')}{(1-\xi^2)m^2M}  \,(\xi G_{T3}-G_{T4}) \right.\nn\\
     &-&\left. 4 \frac{1-\xi}{M}\Big(\kappa_+\,G_{T5} + (1-\xi^2)\frac{t'}{2m}\,G_{T6} + (1-\xi)m\,G_{T7}\Big)
       \right.\nn\\
       &&\left.-2\frac{(1-\xi^2)t'+2\xi\kappa_-M}{mM}\,G_{T8}  \right\}\,, \nn\\
       A^\Delta_{-1+,+-}&=& -\frac{ \sqrt{1-\xi^2}}{4\sqrt{6}}\, \frac{t'}{m^2}\,\left\{  
              \frac{\kappa_-(\kappa_++(1+\xi)M)+(1-\xi^2)t'}{(1-\xi^2)mM}\,G_{T2}\right.\nn\\
               &+& \left. G_{T3} + G_{T4}  
              - 2 (1-\xi)\frac{m}{M}\,G_{T6} - 2\frac{m}{M}\,G_{T8}\right\}\,.
   \label{eq:A-trans}
   \ea 

   Notice that the set of GPDs $G_{Ti}$, defined in \req{eq:trans-GPDs}, is linearly independent: the
   determinant of the $8\times 8$ matrix that relates the GPDs to the matrix elements, $A^\Delta$, is
   non-zero for $t'\neq 0$.

   The matrix elements, $A^\Delta$, are suitable for the calculation of amplitudes for any meson-$\Delta(1232)$
   final state. In particular the $\gamma^*p\to \pi^-\Delta^{++}$ amplitudes are given by the convolutions
   of quark-helicity-flip subprocess amplitudes which are of twist-3 nature, with the matrix elements
   $A^\Delta_{\nu'-\lambda\nu\lambda}$:
   \ba
      {\cal M}_{0\nu'\mu\nu}^{tw3}\= e_0\int_{-1}^1 dx \Big[{\cal H}^{\pi^-}_{0-\mu+} A^\Delta_{\nu'-\nu+} +
        {\cal H}^{\pi^-}_{0+\mu-} A^\Delta_{\nu'+\nu-}\Big]
      \label{eq:tw3-ampl}
   \ea
   Only transversally polarized photons contribute to the twist-3 amplitudes.
   
   It can be checked that
   \be
   A^\Delta_{-\nu'-\lambda'-\nu-\lambda}\= (-1)^{\nu'-\lambda'-\nu+\lambda}\,A^\Delta_{\nu'\lambda'\nu\lambda}\,.
   \ee
   As we already mentioned the analogous property of the  helicity amplitudes (see \req{eq:parity-M})
   holds as a consequence of parity invariance.
   Inspection of \req{eq:A-long} and \req{eq:A-trans} furthermore reveals that for $t'\to 0$
   the matrix elements behave as
   \be
   A^\Delta_{\nu'\lambda'\nu\lambda} \sim \sqrt{-t'}^{|\nu'-\lambda'-\nu+\lambda|}\,.
   \label{eq:t-0-A}
   \ee
   Also this property of the matrix elements is shared by the center-of-mass helicity amplitudes \ci{wang66}.
   
    For $t\to 0$ the subprocess helicity amplitudes behave as \ci{wang66}~\footnote{
      Since the quarks are considered as massless particles $t_0$ for the subprocess is zero.}
   \be
      {\cal H}_{0\lambda'\mu\lambda} \sim \sqrt{-t}^{|-\lambda'-\mu +\lambda|}
   \label{eq:H-t-0}
   \ee
   This property of helicity amplitudes is a consequence of angular momentum conservation.
   Strictly speaking it reflects the conservation of the 3-component of the spin in the collinear situation
   at $t=0$. In the generalized Bjorken regime where $-t\ll Q^2$, the dominant contribution comes from
   subprocess amplitudes evaluated at $t=0$. Hence, these subprocess amplitudes must be helicity non-flip
   ones implying
   \be
   -\mu \= \lambda' - \lambda
   \ee
   Evidently, in this case \req{eq:t-0-A} provides the correct $t'$-dependence of the helicity
   amplitudes for $t'\to 0$ \ci{wang66}
   \be
      {\cal M}_{0\nu'\mu\nu} \sim \sqrt{-t'}^{|-\nu'-\mu+\nu|}\,.
   \label{eq:t-0-M}
      \ee

    Since from \req{eq:H-t-0} follows that for $t\to 0$
      \ba
         {\cal H}^\pi_{0-++}&=&- {\cal H}^\pi_{0+--} \propto {\rm const} \nn\\
         {\cal H}^\pi_{0--+}&=&\phantom{-} {\cal H}^\pi_{0++-} \propto t
         \ea
   the amplitudes \req{eq:tw3-ampl} simplify - one of the two terms in this relation is zero.
   The  $\gamma^*p\to \pi^-\Delta^{++}$ helicity amplitudes for transversally polarized photons read
   \ba
    {\cal M}_{03++}^{tw3}&=& -\frac{e_0}{4\sqrt{2}}\sqrt{\frac{1+\xi}{1-\xi}}\,\frac{t'}{m^2}\,
             \Big[\frac{\kappa_-}{m}\, \langle G_{T2}^{(3)}\rangle + (1+\xi)\,(\langle G_{T3}^{(3)}\rangle -
              \langle G_{T4}^{(3)}\rangle )\Big]\,, \nn\\
    {\cal M}_{03-+}^{tw3}&=&- \frac{e_0}{4\sqrt{2}}\sqrt{\frac{1+\xi}{1-\xi}}\,
             \Big[\,\frac{t'}{m^2}\Big( \frac{\kappa_-}{m} \langle G_{T2}^{(3)}\rangle  
               + (1-\xi)(\langle G_{T3}^{(3)}\rangle +\langle G_{T4}^{(3)}\rangle) \Big)\nn\\
               &&\hspace*{0.2\tw} + 4(1-\xi) \langle G_{T7}^{(3)}\rangle
                      - 4 \frac{\xi}{1+\xi}\frac{\kappa_-}{m}\langle G_{T8}^{(3)}\rangle \Big]\,,\nn\\ 
   {\cal M}_{0-3++}^{tw3}&=&-\frac{e_0}{4\sqrt{2}}\frac{\sqrt{-t'}}{m} 
          \Big[-4 (1+\xi)\,G_{T1} + (1+\xi)\frac{t'}{m^2}\,\langle G_{T2}^{(3)}\rangle  \nn\\
            &&\hspace*{0.2\tw}  + 2\frac{\kappa_+}{m}\,\frac{\xi \langle G_{T3}^{(3)}\rangle
                                  - \langle G_{T4}^{(3)}\rangle}{1-\xi}
                                  - 4\xi \langle G_{T8}^{(3)}\rangle \Big]\,, \nn\\
    {\cal M}_{0-3-+}^{tw3}&=&\frac{e_0}{4\sqrt{2}}\frac{(-t')^{3/2}}{m^3} (1+\xi)               
                         \,\langle G_{T2}^{(3)}\rangle\,, \nn\\
    {\cal M}_{01++}^{tw3}&=& \frac{e_0}{2\sqrt{6}}\frac{\sqrt{-t'}}{m}\frac1{1-\xi} \left\{\phantom{\frac12}
           -2(1+\xi)\,\langle G_{T1}^{(3)}\rangle  \right.\nn\\
          &+&\left.  \frac{\kappa_-^2\kappa_++(1-\xi^2)t'(\kappa_-+(1+\xi)M)}{2(1+\xi)m^2M}\,
                                          \langle G_{T2}^{(3)}\rangle \right.\nn\\
          &+&\left.  \frac{\kappa_+\kappa_- + (1-\xi^2)t'}{2mM}\,\big(\langle G_{T3}^{(3)}\rangle
                                          -\langle G_{T4}^{(3)}\rangle \big) 
          +  \frac{\kappa_+}{m}\,\big(\xi \langle G_{T3}^{(3)}\rangle -\langle G_{T4}^{(3)}\rangle\big) \right.\nn\\
                     &-&\left.  (1-\xi^2)(1-\xi)\frac{m}{M}\,\langle G_{T5}^{(3)}\rangle \right.\nn\\
                     &-&\left. (1-\xi)^2 \frac{\kappa_-}{M}\,\langle G_{T6}^{(3)}\rangle 
                               - (1-\xi)\frac{\kappa_-}{M}\,\langle G_{T8}^{(3)}\rangle \right\}\,, \nn\\
          {\cal M}_{01-+}^{tw3}&=& \frac{e_0}{4\sqrt{6}}\frac{\sqrt{-t'}}{m} 
                \left\{  \frac{\kappa_-^2\kappa_++(1-\xi^2)t'(\kappa_-+(1+\xi)M)}{(1-\xi^2)m^2M}\,
                             \langle G_{T2}^{(3)}\rangle \right.\nn\\
                &+&\left. \frac{\kappa_-\kappa_+ + (1-\xi^2)t'}{(1+\xi)mM}\, \big(\langle G_{T3}^{(3)}\rangle
                             +\langle G_{T4}^{(3)}\rangle\big)
                             + 2(1-\xi)^2\frac{m}{M}\, \langle G_{T5}^{(3)}\rangle          \right.\nn\\
          &-&\left.  2(1-\xi)\Big(\frac{\kappa_-}{M}\,\langle G_{T6}^{(3)}\rangle
                             - 2\frac{m}{M}\, \langle G_{T7}^{(3)}\rangle\Big)
                             - 2\frac{\kappa_-+2\xi M}{M}\,\langle G_{T8}^{(3)}\rangle\right\}\,,  \nn\\
      {\cal M}_{0-1++}^{tw3}&=&\frac{e_0}{4\sqrt{6}}\frac1{\sqrt{1-\xi^2}}  \left\{
                        - 4\frac{\kappa_-\kappa_++(1-\xi^2)t'}{mM}\,\langle G_{T1}^{(3)}\rangle \right.\nn\\
                                   &+& \left. \frac{t'}{m^2} \frac{\kappa_-(\kappa_++(1+\xi)M)
                                     + (1-\xi^2)t'}{mM}\,\langle G_{T2}^{(3)}\rangle \right.\nn\\
          &+&\left. (1+\xi)^2\frac{t'}{m^2}\,\big(\langle G_{T3}^{(3)}\rangle-\langle G_{T4}^{(3)}\rangle\big)
                        \right.\nn\\
                    &+&\left. \frac{2\kappa_+( \kappa_-\kappa_+ +(1-\xi^2)t')}{(1-\xi^2)m^2M}\,
                                 \big(\xi \langle G_{T3}^{(3)}\rangle -\langle G_{T4}^{(3)}\rangle\big)\right.\nn\\
                  &-&\left. 4(1-\xi)\frac{\kappa_+}{m}\, \langle G_{T5}^{(3)}\rangle
                                   - 2(1-\xi)(1-\xi^2)\frac{t'}{Mm}\,\langle G_{T6}^{(3)}\rangle \right. \nn\\
                            &-&\left. 4(1-\xi)^2\frac{m}{M}\,\langle G_{T7}^{(3)}\rangle
                            - 2\frac{2\xi\kappa_-m +(1-\xi^2)t'}{mM}\,\langle G_{T8}^{(3)}\rangle \right\}\,, \nn\\
         {\cal M}_{0-1-+}^{tw3}&=&\frac{e_0}{4\sqrt{6}}\frac1{\sqrt{1-\xi^2}}\frac{t'}{m^2} 
                \left\{ \frac{\kappa_-(\kappa_++(1+\xi)M) + (1-\xi^2)t'}{mM}\,\langle G_{T2}^{(3)}\rangle \right.\nn\\
          &+&\left. (1-\xi^2) \big(\langle G_{T3}^{(3)}\rangle + \langle G_{T4}^{(3)}\rangle \big) \right.\nn\\
                &-&\left. 2 (1-\xi^2)\frac{m}{M}\, \big( (1-\xi)\,\langle G_{T6}^{(3)}\rangle
                + \langle G_{T8}^{(3)}\rangle \big) \right\}\,,
          \label{eq:tw3-amplitudes}      
          \ea
          where
          \be
          \langle G_{Ti}^{(3)} \rangle \= \int_{-1}^1 dx {\cal H}^{\pi^-}_{0-++} G_{Ti}^{(3)}
          \label{eq:tw3-convolutions}
          \ee
        
   %%%%%%%%%%%%%%%%%%%%%%%%%%%%%%%%%%%%%%%%%%%%%%%%%%%%%%%%%%%%%%%%%%%%%%%%%%%%%%%%%%%%%%%%%%%
   \section{The pion-pole contribution}
   %%%%%%%%%%%%%%%%%%%%%%%%%%%%%%%%%%%%%%%%%%%%%%%%%%%%%%%%%%%%%%%%%%%%%%%%%%%%%%%%%%%%%%%%%%%%
   \begin{figure}
      \centering
  \includegraphics[width=0.35\tw]{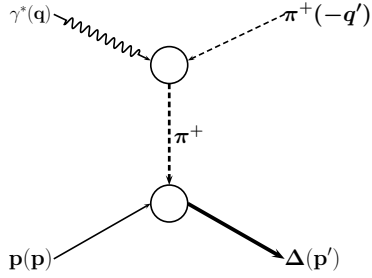}
      \caption{The pion pole contribution. }
      \label{fig:pion-pole}
   \end{figure}
   The pion pole contribution, treated as a one-boson exchange term, reads
\be
   {\cal M}_{0\nu'\mu\nu}^{\rm pole}\=e_0\frac{\varrho_{\pi\Delta}}{t-m_\pi^2}\bar{u}_\delta(p',\nu')
   \frac{\Delta^\delta}{m} u(p,\nu)\,(q - 2q')_\rho\eps^\rho(\mu)\,,
   \label{eq:pion-pole}
   \ee
   where
   \be
   \varrho_{\pi\Delta}\=\sqrt{2} g_{\pi\Delta^{++}p}F_{\pi\Delta p}(t) F_\pi(Q^2)\,.
   \label{eq:rho}
   \ee
   As usual a possible $t$-dependence of the pion form factor, $F_\pi$, is ignored (remember $-t\ll Q^2$
   in the generalized Bjorken regime). The coupling of the pion to
   the proton and the $\Delta^{++}$ is described by a coupling constant, $g_{\pi\Delta^{++}p}$, and a $t$-dependent
   form factor, $F_{\pi\Delta p}$.
   Explicitly, the pole contribution to the $\gamma^*p\to \pi^-\Delta^{++}$ amplitudes reads:
   \ba
      {\cal M}_{030+}^{\rm pole}&=& -\frac{e_0}{\sqrt{2}} \frac{\varrho_{\pi\Delta}}{t - m^2_\pi}
      \frac{\sqrt{-t'}}{m}\,Q\,\frac{\kappa_+}{1-\xi}\,, \nn\\
      {\cal M}_{0-30+}^{\rm pole}&=& \frac{e_0}{\sqrt{2}} \frac{\varrho_{\pi\Delta}}{t - m^2_\pi}
          \frac{t'}{m}\, Q\,\sqrt{\frac{1+\xi}{1-\xi}}\,, \nn\\
      {\cal M}_{03\pm +}^{\rm pole}&=& \mp \frac{e_0}{2} \frac{\varrho_{\pi\Delta}}{t - m^2_\pi}
           \, \sqrt{\frac{1+\xi}{1-\xi}} \,\frac{t'}{m}\,\kappa_+\,, \nn\\
      {\cal M}_{0-3\pm +}^{\rm pole}&=& \pm \frac{e_0}{2}\frac{\varrho_{\pi\Delta}}{t - m^2_\pi}
           \,\frac{(-t')^{3/2}}{m}\,(1+\xi)\,,  \nn\\
      {\cal M}_{010+}^{\rm pole} &=& -\frac{e_0}{\sqrt{6}}\,\frac{\varrho_{\pi\Delta}}{t - m^2_\pi}\,
               \frac{Q}{mM}\, 
               \frac{(1-\xi^2) t'\Big(\kappa_+ + (1+\xi) M \Big)+ \kappa_-\kappa_+^2}{(1-\xi^2)^{3/2}}\,, \nn\\  
      {\cal M}_{0-10+}^{\rm pole} &=& \frac{e_0}{\sqrt{6}} \frac{\varrho_{\pi\Delta}}{t - m^2_\pi}\,
           \sqrt{-t'}\,\frac{Q}{mM} 
               \,\frac{ (1-\xi^2) t' + \kappa_+(\kappa_- + (1+\xi)M)}{1-\xi^2}\,, \nn\\
        {\cal M}_{01\pm +}^{\rm pole} &=&\pm \frac{e_0}{2\sqrt{3}}\,\frac{\varrho_{\pi\Delta}}{t - m^2_\pi}
        \, \frac{\sqrt{-t'}}{mM}\, \frac{(1-\xi^2)t'(\kappa_++(1+\xi)M) + \kappa_-\kappa_+^2}{1-\xi^2}\,,\nn\\
         {\cal M}_{0-1\pm +}^{\rm pole} &=&\pm \frac{e_0}{2\sqrt{3}}\,\frac{\varrho_{\pi\Delta}}{t - m^2_\pi}\,
         \frac{t'}{mM}\, \frac{ (1-\xi^2)t' + \kappa_+(\kappa_-+(1+\xi)M)}{\sqrt{1-\xi^2}} \Big]\,. 
        \label{eq:pole-contributions}
           \ea
      Taking into account the pion form factor which is hidden in $\varrho_{\pi\Delta}$ (see \req{eq:rho}),
      with its $1/Q^2$ behavior the pion-pole contribution to the longitudinal amplitudes behaves as $1/Q$
      asymptotically whereas the transverse ones drop as $1/Q^2$. Thus, in this respect, the pion-pole
      contribution to the $\gamma^*p\to \pi^-\Delta$ amplitudes behaves as that one to the
      $\gamma^* n\to \pi^-p$ amplitudes.
      There are corrections to the amplitudes given in \req{eq:pole-contributions} suppressed by
      $1/Q^2$ which are extremely complicated. They can be neglected even for the 
      longitudinal amplitudes since they are still suppressed by $1/Q$ compared to the transverse amplitudes.
      
      The amplitudes satisfy the familiar symmetry relation for pion (unnatural parity) exchange
     \be
              {\cal M}^{\rm pole}_{0\nu'\mp \nu}\= - {\cal M}^{\rm pole}_{0\nu'\pm \nu}
             \label{eq:U-symmetry}
      \ee
      This is the same relation as for $\gamma^*p\to \pi N$ because both processes possess the same upper
      vertex. The relation \req{eq:U-symmetry} follows from the dynamics and forces the following amplitudes
      \be
           {\cal M}^{\rm pole}_{0-1++}, \quad  {\cal M}^{\rm pole}_{03-+}, \quad  {\cal M}^{\rm pole}_{0-3++}
      \ee
      to vanish by a factor of $t'$ faster to zero for $t'\to 0$ then is forced by angular
      momentum conservation, see \req{eq:t-0-M}.

     For longitudinally polarized photons the pion-pole contribution \req{eq:pion-pole} can be cast into the form
    \be
   {\cal M}^{\rm pole}_{0\nu'0\nu}\= \frac{e_0}{2P^+} \bar{u}(p',\nu')_\delta\frac{\Delta^\delta \Delta^+}{m^2}u(p,\nu)\,
       \Big[\frac{\varrho_{\pi\Delta}}{t-m_\pi^2} q'\cdot\eps(0)\frac{2m}{\xi}\Big]
    \ee
If one is interested in the leading-twist contribution which dominates for $Q^2\to\infty$, than the
comparison with \req{eq:odd-pD-GPDs} and \req{eq:long-amplitudes} reveals that the pion-pole contributes to
the GPD $\tilde{G_4}$ asymptotically. The term in brackets is its convolution with the leading-twist
subprocess amplitude
    \be
       {\cal H}^{\pi^-}_{0+0+} \= 2\frac{Q F_\pi^{\rm pert.}(Q^2)}{f_\pi\langle 1/\tau\rangle_\pi}
                                   \Big[\frac{e_d}{x-\xi +i\eps} + \frac{e_u}{x+\xi-i\eps}\Big]
    \ee
    where $f_\pi=0.131\,\gev$ is the pion decay constant, $\langle 1/\tau \rangle_\pi$ denotes the $1/\tau$ moment
    of the leading-twist pion distribution amplitude, $\Phi_\pi$, and $e_a$ is the charge of the flavor-$a$ quark
    in units of the positron charge, $e_0$. Finally, $F_\pi^{\rm pert.}$ is the leading-twist, leading-order result
    for the electromagnetic form factor of the pion \ci{farrar79,rad79,brodsky79}.
    This result for the convolution can be achieved if in analogy to the case of
    $\gamma^*p \to \pi^+ n$ \ci{goeke99,man99} the GPD is
    \be
    \tilde{G}_4^{\rm pole}\=\tilde{G}_{4 {\rm pole}}^u -\tilde{G}_{4 {\rm pole}}^d\=\Theta(|x|\leq \xi)\,
                \Phi_\pi(\frac{x+\xi}{2\xi})\,
                \frac{ m f_\pi}{\sqrt{2}\xi}\frac{g_{\pi\Delta N}F_{\pi \Delta N}(t)}{t-m_\pi^2}
    \ee

    The difference between the leading-twist contribution from the pion pole and its full contribution
    \req{eq:pion-pole} is the replacement of $F_\pi^{\rm pert.}$ by the full form factor as extracted from
    the data on the longitudinal $\gamma^*p\to \pi^+n$ cross section \ci{volmer01}. Since $F_\pi^{\rm pert.}$ is
    substantially smaller than its experimental value a leading-twist analysis of pion electroproduction
    within the handbag approach evidently fails. In \ci{GK5} the pion-pole term is, therefore, treated
    as an one-bose-exchange contribution in the analysis of exclusive $\pi^+$ data \ci{hermes08} and
    reasonable agreement with experiment is obtained, see also \ci{favart16}. We are going to apply the
    same procedure to the process of interest here, $\gamma^*p\to \pi^-\Delta^{++}$.

    In the evaluation of observables for $\gamma^*p\to \pi^-\Delta^{++}$ we will use the following parameterization
    of the from factors \ci{GK5,GK6}
    \be
    F_\pi\=\frac1{1+2.0\, \gev^{-2} Q^2}\,, \qquad F_{\pi\Delta p}\=\frac{\Lambda_N^2-m_\pi^2}{\Lambda_N^2-t}
    \ee
    with $\Lambda_N=0.44\,\gev$. For the $\pi\Delta^{++} p$ coupling constant we take the value \ci{ellis97}
    \be
    g_{\pi\Delta^{++}p}\=14.8\,.
    \label{eq:coupling-constant}
    \ee

%%%%%%%%%%%%%%%%%%%%%%%%%%%%%%%%%%%%%%%%%%%%%%%%%%%%%%%%%%%%%%%%%%%%%%%%%%%%%%%%%%%%%%%%%%%%%%%%%%%%%%%%
    \section{GPDs in the large-$N_C$ limit}
\begin{figure}
      \centering
      \includegraphics[width=0.4\tw]{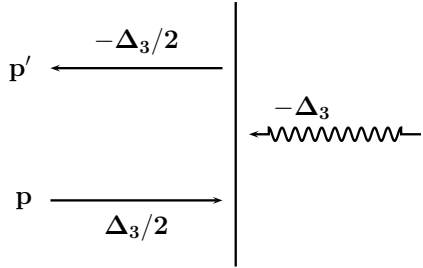}
      \caption{The Breit frame.}
      \label{fig:Breit-frame}
      \end{figure}

    %%%%%%%%%%%%%%%%%%%%%%%%%%%%%%%%%%%%%%%%%%%%%%%%%%%%%%%%%%%%%%%%%%%%%%%%%%%%%%%%%%%%%%%%%%%%%%%%%%%%%%%
    As we have seen there are 12 GPDs contributing to the exclusive electroproduction of  $\pi^-\Delta^{++}$.
    All of them are unknown functions and, at present, there are no relevant data to fix them.  
    In order to be able to make predictions, or better estimates, we therefore take recourse to a theoretical approach,
    namely the large-$N_C$ limit in which the $p-\Delta^+$ GPDs are related to the proton-proton ones
    \ci{belitsky,frankfurt99}. In this limit the nucleon and the $\Delta(1232)$ are different rotational
      excitations of the same classical object, the chiral soliton, and are degenerated in mass.
      This leads to relations between the matrix elements of the quark-field operators, \req{eq:Opp} or
      \req{eq:Omp}, in the isovector combination \ci{belitsky}
      \be
      \sqrt{2} \langle p_{\uparrow (\downarrow)}|{\cal O}^{(3)}|p_\uparrow\rangle \=
      \langle \Delta^+_{\uparrow(\downarrow)}|{\cal O}^{(3)}|p_\uparrow\rangle
      \label{eq:soliton}
      \ee
      The arrows denote spin states. From \req{eq:soliton} relations between the $p-\Delta^+$ GPDs and
      the proton-proton ones follow. Subsequently, isospin symmetry \req{eq:isospin} links the
      $p-\Delta^+$ GPDs to the $p-\Delta^{++}$ ones. Since the baryon mass, $M_C$, is large, of order
      $N_C$, one has to make a non-relativistic reduction of the Dirac bilinears appearing in
      \req{eq:odd-pD-GPDs} and \req{eq:trans-GPDs}. This can be done consistently in a Breit frame,
      see Fig. \ref{fig:Breit-frame}. In this frame there is no energy transfer but only a momentum
      transfer, $\Delta_3$. For the $p-p$ GPDs we use those defined in \ci{diehl01}. The matrix elements
      in \req{eq:soliton} are then calculated in the Breit frame and expanded in powers of $\Delta_3/M_C$.
      The coefficients of the various powers of $\Delta_3/M_C$ for the two matrix elements are equated which
      provides the desired relations between the proton-proton GPDs and the $p-\Delta^+$ ones.
      In combination with the isospin relation \req{eq:isospin} one obtains
      for the odd-parity $p-\Delta^{++}$ GPDs with this method \ci{belitsky,frankfurt99}
      \ba
      \tilde{G}_3^{(3)}&=&\frac32 \big( \tilde{H}^u-\tilde{H}^d\big)\,,  \nn\\
      \tilde{G}_4^{(3)}&=& \frac38 \big(\tilde{E}^u-\tilde{E}^d\big)\,.
      \label{eq:tilde-GPDs-large-NC}
      \ea
      where, for each GPD, only the leading contribution in $N_C$ is taken into account.
      It can be shown that \ci{weiss16}
      \be
      \tilde{H}^u-\tilde{H}^d\sim N_C^2\,, \qquad \tilde{E}^u-\tilde{E}^d\sim N_C^4\,.
      \ee
      The respective opposite flavor combinations are suppressed by $1/N_C$.

      The corresponding application of this method to the matrix elements of  the operator \req{eq:Omp}
      leads to the relation
      \be
      G^{(3)}_{T5} + \frac12\,G^{(3)}_{T7}\=-\frac32\Big( H^u_T - H^d_T\Big)
      \label{eq:trans-GPDs-large-NC}
      \ee
      between the $p-\Delta^{++}$ and the $p-p$ GPDs for the leading term of the $\Delta_3/M_C$ expansion.
      The higher-order terms relate even more complicated combinations of $p-\Delta$ GPDs to either the
      $p-p$ transversity GPD $\tilde{E}_T$ or $E_T$ which are unknown as yet and usually neglected in
      applications of the handbag approach \ci{GK5,GK6,liuti}. The GPDs $\tilde{H}_T$ and
      $\bar{E}_T=2\tilde{H}_T+E_T$ do not contribute to the matrix elements
      $\langle p_\downarrow|{\cal O}^{(3)}_{\mp\pm}| p_\uparrow\rangle$ evaluated in the Breit frame.
      This is in accordance with the large-$N_C$ result, that the isovector combinations of these
      GPDs are suppressed compared to the sum of the corresponding $u$- and $d$-quark GPDs \ci{weiss16}.
      Hence, the large-$N_C$ results for the transversity GPDs are of no help
      without additional assumptions. In the following we are going to probe two assumptions: all $p-\Delta$
      transversity GPDs are zero except of either $G_{T5}$ or $G_{T7}$ and apply \req{eq:trans-GPDs-large-NC}
      to the non-zero one.
  %%%%%%%%%%%%%%%%%%%%%%%%%%%%%%%%%%%%%%%%%%%%%%%%%%%%%%%%%%%%%%%%%%%%%%%%%%%%%%%%%%%%%%%%%%%%
\section{The subprocess amplitudes and the $p-p$ GPDs}
%%%%%%%%%%%%%%%%%%%%%%%%%%%%%%%%%%%%%%%%%%%%%%%%%%%%%%%%%%%%%%%%%%%%%%%%%%%%%%%%%%%%%%%%%%%
Due to the large-$N_C$ results \req{eq:tilde-GPDs-large-NC} and \req{eq:trans-GPDs-large-NC} and because
the processes $\gamma^*p\to \pi^-\Delta^{++}$ and $\gamma^*n\to \pi^-p$ have the same hard subprocesses, namely
$\gamma^*d\to \pi^-u$, the convolutions \req{eq:tw2-convolutions} and \req{eq:tw3-convolutions} are the same
as those of the $p-p$ GPDs $\tilde{H}^{(3)}$, $\tilde{E}^{(3)}$ and $H_T^{(3)}$ up to numerical prefactors.
The calculation of the latter has been performed in previous work \ci{GK5,GK6,GK3} within the generalized
handbag approach and is described in great detail therein~\footnote{
  For the process $\gamma^*n\to \pi^-p$ the $n-p$ transition GPDs occur which, by flavor symmetry
  \ci{frankfurt,man99}, are related to the isovector combination of the proton-proton GPDs
  $K^{ud}_{n\to p}=K^u-K^d$ for any GPD $K=H, E, \tilde{H}, \ldots$.}.
Therefore, only the basics facts will be sketched in this section.

The main idea of the generalized handbag approach is to keep the quark transverse momentum, $k_\perp$,
in the subprocess while the emission and reabsorption of the partons from the baryons are still treated
collinear to the baryon momenta. The subprocess amplitudes (at $t=0$) read
\ba
{\cal H}^\pi_{0\lambda',\mu\lambda} &=& \int d\tau d^2b\, \hat{\Psi}_{\pi,-\lambda'\lambda}
                                                              (\tau,-{\bf b},\mu_F)\,
                      \hat{F}^\pi_{0\lambda'\mu\lambda}(x,\xi,\tau,Q^2,{\bf b},\mu_R)\nn\\
                  &&\times           \als(\mu_R)\exp{[-S(\tau,{\bf b},Q^2,\mu_F,\mu_R]}
\label{eq:subprocess}
\ea
in the impact parameter space; ${\bf b}$ is canonically conjugated to the quark transverse
momenta. The Sudakov factor, $S$, has been calculated by Botts and Sterman \ci{botts} in
next-to-leading-log approximation using resummation techniques and having recourse to the
renormalization group. It takes into account the gluon radiation resulting from the 
separation of color charges which is a consequence of the quark transverse momenta.
The properties of the Sudakov factor force the following choice of the factorization
scale: $\mu_F=1/b$. The renormalization scale is taken to be the largest mass scale
appearing in the subprocess, i.e. $\mu_R={\rm max}(\tau Q,(1-\tau) Q,1/b)$ ($\tau$ is the
momentum fraction of the quark entering the meson). The strong coupling constant, $\als$, is
evaluated from four flavors and $\LQCD=0.181\,\gev$.
The inclusion of the quark transverse momenta and the Sudakov factor has two advantages -
firstly the magnitude of the subprocess amplitudes are somewhat reduced as compared to a
collinear calculation which leads to a better agreement with experiment (see e.g.\ \ci{GK6})
and, secondly, the infrared singularity occurring in a collinear calculation of the twist-3
subprocess amplitude is regularized.

For the twist-2 and twist-3 hard scattering kernels, $F^\pi$, evaluated to lowest order of 
perturbative QCD, it is referred to Refs.\ \ci{GK5,GK3}. The last object to be explained 
is $\hat{\Psi}_{\pi,\lambda'\lambda}$. It represents the Fourier transform of a soft pion 
light-cone wave function. For twist 2 one has $\lambda'=\lambda$ and the distribution
amplitude associated to $\hat{\Psi}$, is the familiar twist-2 one. On the other hand, for
twist 3 one has $\lambda'=-\lambda$ and a twist-3 wave function is required. The wave functions
are parameterized as Gaussians in $k_\perp$ \ci{GK5}:
\ba
\Psi_{\pi, -+}&=& 8\pi^2 \frac{f_\pi}{\sqrt{2N_C}}\frac{a_\pi^2}{\tau\taub}\Phi_\pi(\tau)\,
               \exp{[-a_\pi^2k_\perp^2/(\tau\taub)]}  \nn\\
\Psi_{\pi,++}&=&\frac{16\pi^{3/2}}{\sqrt{2N_C}} f_\pi a_P^3 k_\perp\,\exp{[-a_P^2k_\perp^2]}
\ea
where $\taub=1-\tau$. For the twist-2 and twist-3 transverse size parameters of the pion we take the
values (see \ci{GK6}) 
\be
a_\pi\=0.859\,\gev^{-1}\,, \qquad a_P\=1.8\,\gev^{-1}\,.
\ee
For the familiar leading-twist pion distribution amplitude, $\Phi_\pi$, we simply take the asymptotic
form $6\tau\taub$. The distribution amplitude associated with the twist-3 wave function, $\Psi_{\pi,++}$,
is $\Phi_P\equiv 1$ as is fixed by the equation of motion if the 3-body twist-3 distribution amplitude
is assumed to be zero \ci{braun90}. This ansatz for the twist-3 wave function makes it clear that the
the corresponding subprocess amplitude is calculated in the Wandzura-Wilczek approximation~\footnote{
              In \ci{KP21} we have calculated the full twist-3 subprocess amplitude, i.e.\ its 2-body 
              as well as its 3-body contribution. The application of this result to hard exclusive
              electroproduction of mesons is in progress.}.

For the numerical studies to be 
presented below, a value of $2.0\,\gev$ is used for $\mu_\pi$ at the initial scale $\mu_0=2\,\gev$. 
Since the current-quark masses decrease with increasing scale $\mu_\pi$ is scale dependent. 
The respective anomalous dimension is $4/\beta_0=12/25$ for four flavors.

We will make use of the $p-p$ GPDs determined in \ci{GK5,GK6,GK3}. The idea is to parameterize the
zero-skewness GPDs. Their products with suitable weight functions are considered as double distributions
from which the full, skewness-dependent GPDs can be calculated \ci{musatov99}.
The zero-skewness GPDs for flavor $a$ are parameterized as
\be
K_i^a(z,\xi=0,t) \= K_i^a(x,\xi=t=0) \exp{[(b_i^a-\alpha_i'{}^a\ln(x))t]}\,.
\label{eq:profile}
\ee
The forward limits, $\xi,t\to 0$, of the GPDs $\widetilde{H}$ and $H_T$ are given by 
the polarized and transversity parton densities, respectively. In order to respect
the Soffer bound the transversity density is parameterized as %\ci{anselmino}
\be
\delta^a\=N^a_{H_T} \sqrt{x}\,(1-x)\,\Big[q^a(x)+\Delta q^a(x)\Big]\,.
\ee
The unpolarized and polarized parton densities are taken from \ci{ABM11} and \ci{DSSV09},
respectively. For the non-pole part of $\tilde{E}$ the forward limit is not accessible in
deep inelastic lepton-nucleon scattering and, hence, unknown. Therefore, it is
parameterized like the PDFs
\be
\tilde{E}^{{\rm n.p.}a}(x,\xi=t=0)\= N_e^a x^{-\alpha_e^a(0)}(1-x)^{\beta_e^a}
\ee
with the additional parameters to be adjusted to the electroproduction data.
The  parameters of the zero-skewness GPDs, compiled in Tab.\ \ref{tab:gpd}, are taken from \ci{GK6}. 
The powers $\beta_e^a$ are set to the following values
\be
\tilde{E}^{\rm n.p.}:  \qquad  \beta_e^u\=\beta_e^d\=5\,.
\ee
Since a flavor-symmetric sea is assumed only the valence-quark GPDs, strictly speaking only their
isovector combinations, are needed.
In this case the convolutions 
\req{eq:pm}
and 
\req{eq:tw3-convolutions}
are integrated from  
$-\xi$ to 1.

As is well-known the GPDs evolve with the scale. Since the factorization of the amplitudes into
GPDs and a subprocess is treated collinearly, the GPDs do not know about the impact-parameter
dependence in the subprocess - $b$ is integrated over. Hence, the subprocess factorization scale,
$\mu_F$, does not apply to the GPDs, it refers to the factorization
of the soft meson wave function and the remaining hard part of the subprocess. The scale
of the GPDs is therefore taken as the photon virtuality. 

\begin{table*}[t]
\renewcommand{\arraystretch}{1.4} 
\begin{center}
\begin{tabular}{| c || c | c | c || c | c |}
\hline   
GPD & $\alpha(0)$ & $\alpha^\prime [\gev^{-2}]$ & $b [\gev^{-2}]$ & $N^u$ &
$N^d$ \\[0.2em]  
\hline
$\widetilde{H}$ & - & 0.45  &  0.59  & -  & - \\[0.2em]
$\tilde{E}^{\rm n.p.}$ & 0.48 & 0.45 & 0.9 & 14.0 & 4.0 \\[0.2em]
$H_T$ & - & 0.45 & 0.3 & 1.1 & -0.3 \\[0.2em]
%$\bar{E}_T$& 0.3 & 0.45 & 0.5 & 4.83 & 3.57 \\[0.2em]
\hline
\end{tabular}
\end{center}
\caption{Regge parameters and normalizations of the valence-quark $p-p$ GPDs, quoted at the initial scale
 $\mu_0=2\,\gev$. Lacking parameters indicate that the corresponding parameters
are part of the parton densities.}
\label{tab:gpd}
\renewcommand{\arraystretch}{1.0}   
\end{table*}

%%%%%%%%%%%%%%%%%%%%%%%%%%%%%%%%%%%%%%%%%%%%%%%%%%%%%%%%%%%%%%%%%%%%%%%%%%%%%%%%%%%%%%%%%%%%%%%%%%%%%
\section{Predictions}
%%%%%%%%%%%%%%%%%%%%%%%%%%%%%%%%%%%%%%%%%%%%%%%%%%%%%%%%%%%%%%%%%%%%%%%%%%%%%%%%%%%%%%%%%%%%%%%%%%%%%
The full $\gamma^*p\to \pi^-\Delta^{++}$ amplitudes for longitudinally polarized photons are
\be
   {\cal M}_{0\nu'0\nu} \= {\cal M}^{tw2}_{0\nu'0\nu} + {\cal M}^{\rm pole}_{0\nu'0\nu}
   \ee
   and for transversally polarized photons ($\mu=\pm 1$)
   \be
      {\cal M}_{0\nu'\mu\nu} \= {\cal M}^{tw3}_{0\nu'\mu\nu} + {\cal M}^{\rm pole}_{0\nu'\mu\nu}
      \ee
      For the twist-3 contributions we consider the two scenarios
      \ba
      \phantom{I}I: &\quad& G_{T5}^{(3)}\=-\frac32 H_T^{(3)}\,, \quad G_{Ti}\=0\,,\quad i\neq 5\,, \nn\\
      II: &\quad&  G_{T7}^{(3)}\=- 3 H_T^{(3)}\,, \quad G_{Ti}\=0\,, \quad i\neq 7\,.
      \ea
      For scenario I the twist-3 amplitudes \req{eq:tw3-amplitudes} simplify to
      \ba
  {\cal M}^{tw3}_{01++}&=&\frac{e_0}{4}\sqrt{\frac32} \frac{\sqrt{-t'}}{M} (1-\xi^2) \langle H_T^{(3)}\rangle\,,
           \nn\\
  {\cal M}^{tw3}_{01-+}&=&-\frac{e_0}{4}\sqrt{\frac32} \frac{\sqrt{-t'}}{M} (1-\xi)^2 \langle H_T^{(3)}\rangle\,,
    \nn\\
  {\cal M}^{tw3}_{0-1++}&=& \frac{e_0}{2}\sqrt{\frac32} \sqrt{\frac{1-\xi}{1+\xi}} \frac{\kappa_+}{m}
  \langle H_T^{(3)}\rangle\,,
     \label{eq:scenario-I}
    \ea
    and all other twist-3 helicity amplitudes are zero. For scenario II the non-zero twist-3 amplitudes are
    \ba
    {\cal M}^{tw3}_{03-+}&=& 3 \frac{e_0}{\sqrt{2}} \sqrt{1-\xi^2} \langle H_T^{(3)}\rangle\,, \nn\\
    {\cal M}^{tw3}_{01-+}&=& - e_0 \sqrt{\frac32} \frac{\sqrt{-t'}}{M}(1-\xi)  \langle H_T^{(3)}\rangle\,, \nn\\
    {\cal M}^{tw3}_{0-1++}&=&  e_0 \sqrt{\frac32} \sqrt{\frac{1-\xi}{1+\xi}}(1-\xi)
                                  \frac{m}{M} \langle H_T^{(3)}\rangle\,.
    \label{eq:scenario-II}
    \ea
    For both the scenarios the helicity non-flip amplitudes dominate at small $-t$, ${\cal M}^{tw3}_{0-1++}$
    for scenario I and ${\cal M}^{tw3}_{03-+}$ for scenario II.
    
    The electroproduction cross section for the $\pi^-\Delta^{++}$ final state is defined by
    \ba
\frac{d\sigma^4}{dW^2dQ^2dtd\phi}&=& \frac{\ale(W^2-m^2)}{16\pi^2E_L^2m^2Q^2(1-\veps)}
\left(\frac{d\sigma_T}{dt} + \veps\, \frac{d\sigma_L}{dt} \right. \nn\\
  && \left.    +\, \eps\cos{2\phi}\,\frac{d\sigma_{TT}}{dt}
                    + \sqrt{2\veps(1+\veps)} \cos{\phi}\,\frac{d\sigma_{LT}}{dt}  \right) 
\label{eq:ep-cross-section}
\ea
where $\phi$ is the azimuthal angle between the lepton and the hadron plane, $E_L$ is the beam energy
and $\veps$ is the ratio of the longitudinal and transversal photon fluxes. If a value of the latter
quantity is needed we evaluate it for a beam energy of $10.6\,\gev$ and obtain $\veps=0.77$. The
partial cross sections read
    \ba
\frac{d\sigma_L}{dt} &=& \frac{\sum_{\nu'} |{\cal M}_{0\nu'0+}|^2}{16\pi (W^2-m^2)\sqrt{\Lambda(W^2,-Q^2,m^2)}}\,,\nn\\
\frac{d\sigma_T}{dt} &=& \frac{\sum_{\nu'}\Big[ |{\cal M}_{0\nu'++}|^2 + |{\cal M}_{0\nu'-+}|^2\Big]}
                                                   {32\pi (W^2-m^2)\sqrt{\Lambda(W^2,-Q^2,m^2)}}\,,\nn\\  
\frac{d\sigma_{TT}}{dt} &=& -\frac{\sum_{\nu'} {\rm Re}\Big[ {\cal M}_{0\nu'++}^* {\cal M}_{0\nu'-+}\Big]}
                                             {16\pi (W^2-m^2)\sqrt{\Lambda(W^2,-Q^2,m^2)}}\,,\nn\\
 \frac{d\sigma_{LT}}{dt} &=& -\sqrt{2}\frac{\sum_{\nu'} {\rm Re}\Big[ {\cal M}_{0\nu'0+}^*
                                                 \big({\cal M}_{0\nu'++}-{\cal M}_{0\nu'-+}\big)\Big]}
                                             {32\pi (W^2-m^2)\sqrt{\Lambda(W^2,-Q^2,m^2)}}\,.       
\label{eq:partial-cross-sections}
\ea
The Mandelstam function is defined by
\be
\Lambda(W^2,-Q^2,m^2)\=W^4+Q^4+m^4+2W^2Q^2-2W^2m^2+2Q^2m^2\,.
\ee
Eqs.\ \req{eq:ep-cross-section} and \req{eq:partial-cross-sections} are in agreement with \ci{sapeta05}
but, by definition, the interference cross sections have opposite signs here and in \ci{sapeta05} and the
longitudinal-transverse cross section is larger by a factor $\sqrt{2}$ in \ci{sapeta05} than by us.

\begin{figure}
      \centering
      \includegraphics[width=0.365\tw]{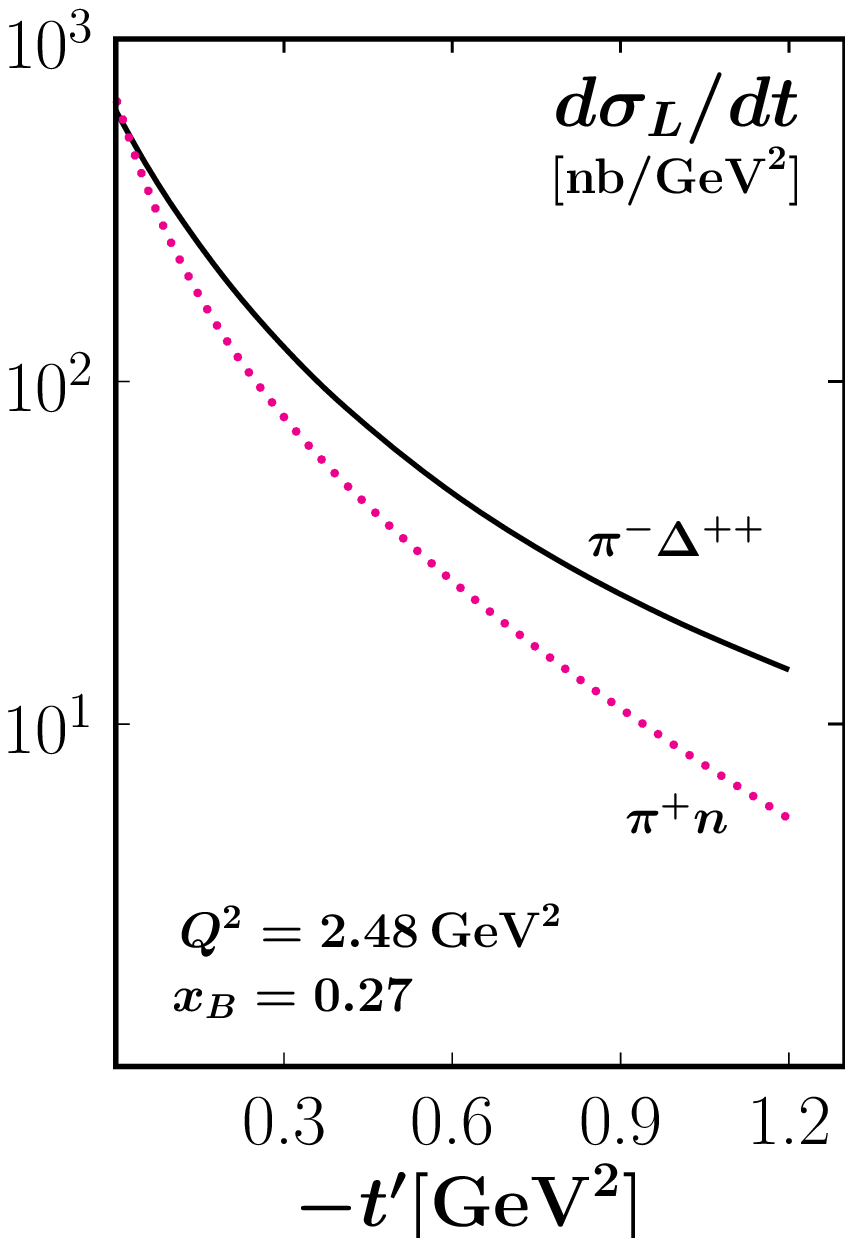} \hspace*{0.18\tw}
      \includegraphics[width=0.365\tw]{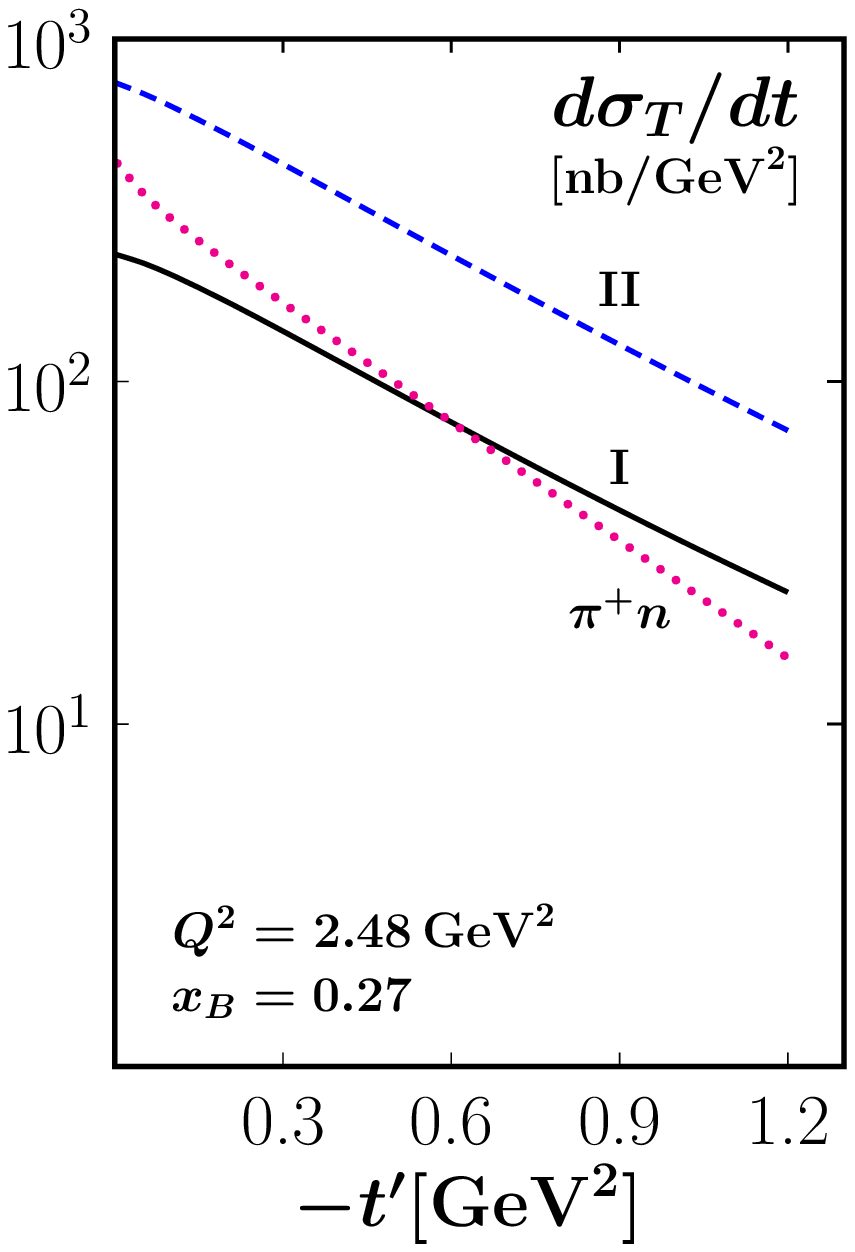}
      \caption{The longitudinal (left) and the transverse (right) cross sections of $\gamma^*p\to \pi^-\Delta^{++}$
        versus $-t'$. The solid (dashed) lines represent the predictions obtained for scenario I (II). For
        comparison the dotted lines are the results for $\gamma^*p\to\pi^+n$ obtained with the same GPDs.}
      \label{fig:sigmaL}
      \vspace*{0.03\tw}
   %\end{figure}
   %\begin{figure}
      \centering
      %\hspace*{-0.09\tw}
      \includegraphics[width=0.365\tw]{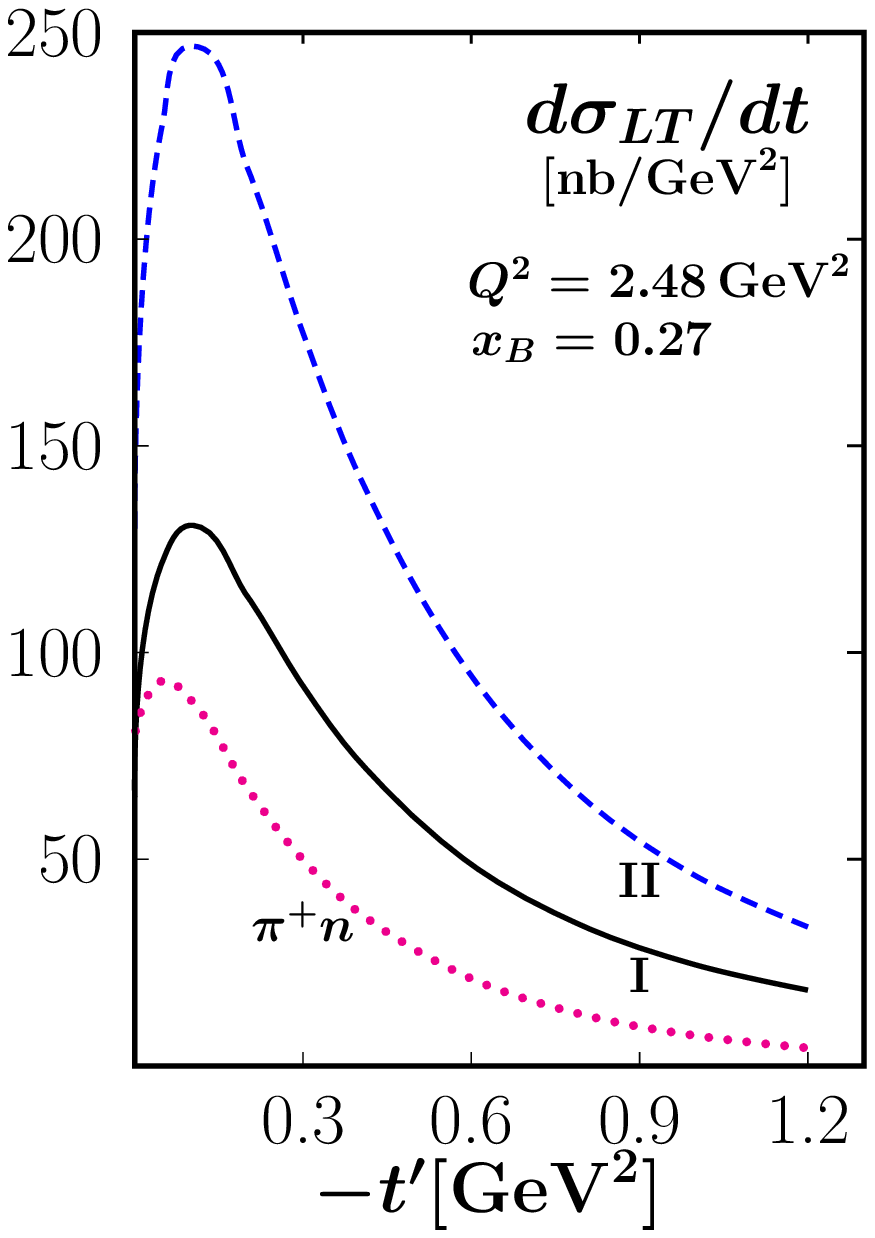} \hspace*{0.18\tw}
      \includegraphics[width=0.355\tw]{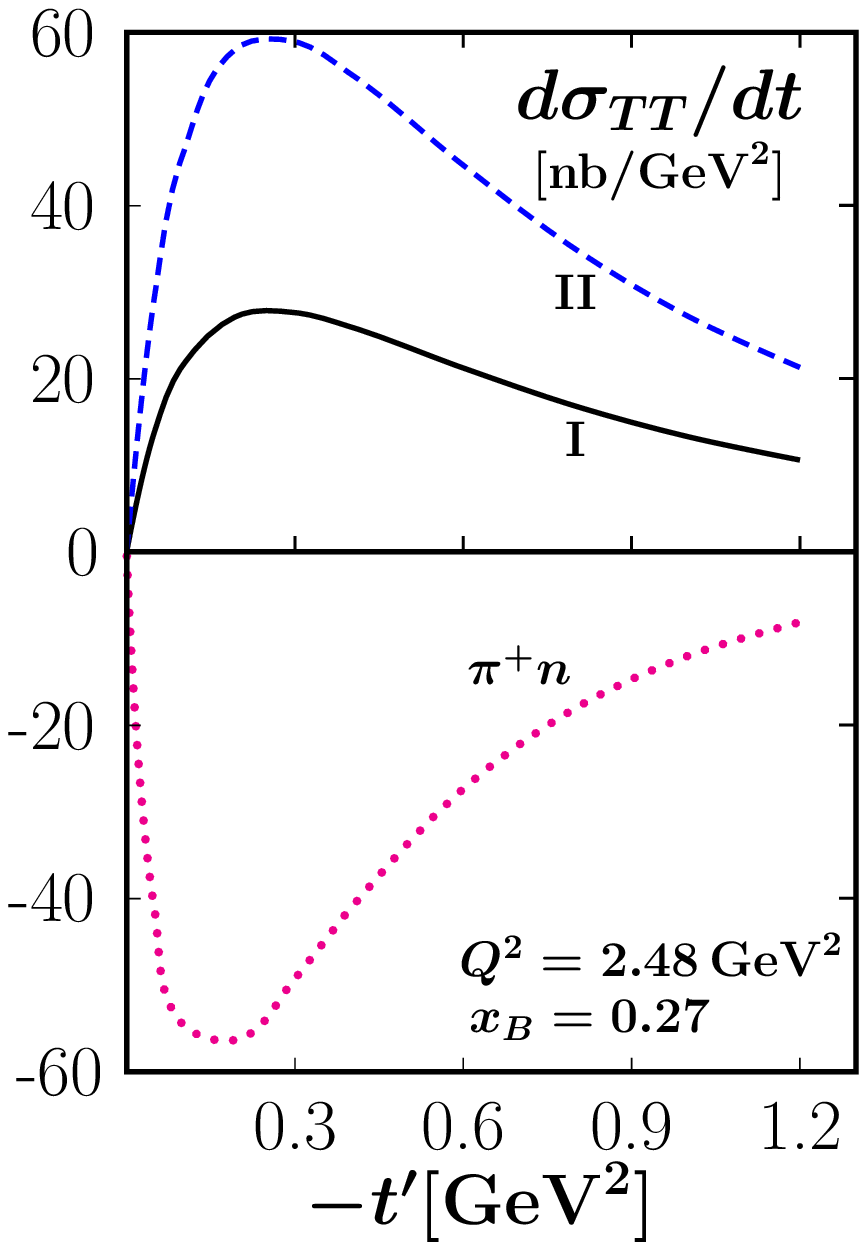}
      \caption{The longitudinal-transverse (left) and the transverse-transverse (right) interference
        cross sections of $\gamma^*p\to \pi^-\Delta^{++}$. For other notations see Fig. \ref{fig:sigmaL}.}
        \label{fig:sigmaTT}
   \end{figure}

In Figs.\ \ref{fig:sigmaL} and \ref{fig:sigmaTT} we display predictions for the partial cross sections for the two
scenarios and compare with the corresponding $\gamma^*p\to \pi^+n$ cross sections. We show the observables as
functions of $t'$ and the reader should be aware of the quite different values of $t_0$ (for the kinematics shown
in the plots $t_0$ is $-0.323\,\gev^2$ and $-0.088\,\gev^2$  for the $\pi^-\Delta^{++}$  and the $\pi^+n$ channel,
respectively).
Therefore, at a given value of $t'$, the  convolutions of a GPD for $\pi^-\Delta^{++}$ and for $\pi^+n$ are
evaluated at different values of $t$. This partially compensates the prefactors in \req{eq:tilde-GPDs-large-NC}
and \req{eq:trans-GPDs-large-NC}.

The longitudinal cross section for which the predictions from the two scenarios fall together, is dominated
by the pion-pole contribution. Since the $\pi\Delta^{++}p$ coupling constant \req{eq:coupling-constant} is
somewhat larger than the familiar pion-nucleon one \ci{compilation} and also the convolution of
$\tilde{G}_{T3}^{(3)}$ is larger than that of $\tilde{H}^{(3)}$ (see \req{eq:tilde-GPDs-large-NC}) it is clear
that the $\pi^-\Delta^{++}$ cross section is larger than the $\pi^+n$ one. For the other partial cross sections
there are substantial differences between the scenarios I and II. One also sees that, at small $-t'$, the
longitudinal cross section is larger than the transverse one as is the case for the $\pi^+n$ channel \ci{GK5}.
The interference cross sections are larger for scenario II than for scenario I. Note that for the $\pi^+n$
channel the transversity GPD $\bar{E}_T$ is taken into account but its contribution is very small \ci{GK6} since,
in agreement with large-$N_C$ results \ci{weiss16}, $\bar{E}_T$ for $u$- and $d$-quarks have the same sign and
about the same size. This is to be contrasted with the $\pi^0p$ channel to which $\bar{E}_T$ contributes in the
combination
\be
\frac1{\sqrt{2}}\,\Big[e_u\bar{E}_T^u-e_d\bar{E}_T^d\Big]
\ee
which is large and provides an important contribution to the $\pi^0p$ cross sections \ci{GK6}. Particularly
interesting is the opposite sign of $d\sigma_{TT}$ for the $\pi^-\Delta^{++}$ and the $\pi^+n$ channels. The
dominant contribution to $d\sigma_{TT}$ comes from the interference of a twist-3 helicity non-flip amplitude,
being proportional to the convolution $\langle H_T \rangle$, and a pion-pole contribution. The $Q^2$ and $W$
dependencies of the $\pi^-\Delta^{++}$ partial cross sections is similar to those of the $\pi^-p$ ones.

The uncertainties of our predictions are very large. First, the large-$N_C$ considerations do not really fix
the transversity GPDs, only a sum of $G_{T5}$ and $G_{T7}$ is related to the proton-proton GPD $H_T$. This forces us
to invent the two scenarios by an additional assumption. 
Furthermore, the quality of the large-$N_C$ results \req{eq:tilde-GPDs-large-NC} is unknown since
there are as yet no $\pi\Delta(1232)$ cross section data available to check this relation. Only the qualitative
results on the relative magnitudes of the $u+d$ and $u-d$ quark combinations of $p-p$ GPDs \ci{weiss16} are
known to be in fair agreement with phenomenology.

   \clearpage

\begin{figure}
      \centering
      \includegraphics[width=0.36\tw]{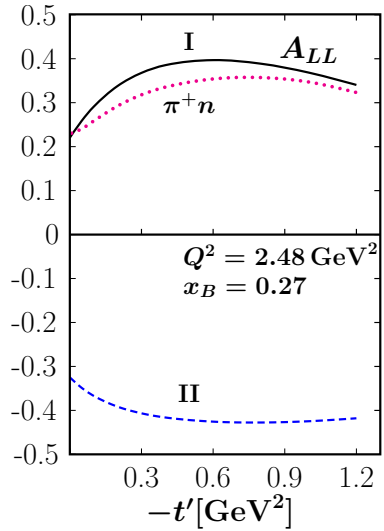}
      \caption{The asymmetry $A_{LL}$ for the $\pi^-\Delta^{++}$ and $\pi^+n$ channels. For other
        notations see \ref{fig:sigmaL}.}
        \label{fig:ALL}
   \end{figure}

For most spin asymmetries the uncertainties are likely even larger than for the cross sections because
they also depend on the imaginary part of products of helicity amplitudes, i.e. from the rather small
differences from amplitude phases. Therefore, we refrain from
showing predictions for asymmetries. An exception is the asymmetry $A_{LL}$ measured with longitudinally
polarized beam and target which, like the transverse cross section, only depends on the absolute values
of the amplitudes for transversally polarized photons:
\be
A_{LL}\= \sqrt{1-\veps^2}\,\frac1{2\sigma_0}\,\sum_{\nu'}\Big[ |{\cal M}_{0\nu'++}|^2 -  |{\cal M}_{0\nu'-+}|^2\Big]
\label{eq:ALL}
\ee
where
\be
\sigma_0\=\sum_{\nu'}\Big[|{\cal M}_{0\nu'++}|^2 + |{\cal M}_{0\nu'-+}|^2 + \veps  |{\cal M}_{0\nu'0+}|^2\big]\,.
\ee
This asymmetry is obtained from a integral upon the electroproduction cross section.
There is a correction due to the fact that the target polarization is defined with respect to the
lepton beam direction in experiment and not with respect to the direction of the virtual photon, see
\ci{sapeta05}. For the kinematics of interest in this work this correction is small and neglected by us.
Predictions of $A_{LL}$ for the two scenarios are shown in Fig.\ \ref{fig:ALL} and compared to this
asymmetry for the $\pi^+n$ channel. The magnitudes of the predictions evaluated from the two scenarios
are close to that one for the $\pi^+n$ channel. However, the predictions for the two scenarios for
$\pi^-\Delta^{++}$ have opposite signs. As an inspection of \req{eq:scenario-I} and \req{eq:scenario-II}
reveals this is understandable: for scenario I
  the dominant helicity non-flip amplitude, $M^{tw3}_{0-1++}$, provides a positive contribution to $A_{LL}$
  (see \req{eq:ALL}) whereas for the second scenario the dominant amplitude, $M^{tw3}_{03-+}$, gives a negative
  contribution to it. The pion pole is unimportant for this asymmetry, most of its contributions cancel
  (note that ${\cal M}^{\rm pole}_{0\nu'++}=-{\cal M}^{\rm pole}_{0\nu'-+}$).

%%%%%%%%%%%%%%%%%%%%%%%%%%%%%%%%%%%%%%%%%%%%%%%%%%%%%%%%%%%%%%%%%%%%%%%%%%%%%%%%%%%%%%%%%%%%%%%%%%%%%%%
\section{Summary}
%%%%%%%%%%%%%%%%%%%%%%%%%%%%%%%%%%%%%%%%%%%%%%%%%%%%%%%%%%%%%%%%%%%%%%%%%%%%%%%%%%%%%%%%%%%%%%%%%%%%%%%
We investigated exclusive electroproduction of $\pi^-\Delta^{++}$  in the generalized Bjorken regime within the
handbag approach. In addition to the known eight helicity non-flip $p-\Delta$ GPDs \ci{belitsky} we defined
 a set of eight transversity GPDs. For both sets of GPDs we calculated the $p-\Delta^{++}$ matrix elements,
$A^\Delta_{\nu'\lambda'\nu\lambda}$, and the helicity amplitudes, ${\cal M}_{0\nu'\mu\nu}$, for 
$\gamma^*p\to \pi^-\Delta^{++}$. We also calculated the pion-pole contribution to this process and
showed that, to leading-twist accuracy, the pion pole contributes to the GPD $\tilde{G}_4$.

In order to generate predictions for the $\pi\Delta^{++}$ partial cross sections we had to take recourse
to the large-$N_C$ limit where a few of the $p-\Delta$ GPDs are related to the diagonal proton-proton ones.
In the large $N_C$ limit the odd-parity helicity non-flip GPDs $\tilde{G}_3$ and $\tilde{G}_4$ in the isovector
combination are related to the corresponding combinations of $\tilde{H}$ and $\tilde{E}$ , respectively
\ci{belitsky,frankfurt99}. For the transversity GPDs we found rather complicated relations. Therefore, we only
utilized the relation between the sum $G_{T5}+G_{T7}/2$ and the $p-p$ GPD $H_T$ obtained from the leading-order
term in the $\Delta_3/M_C$ expansion. We were forced to make the additional assumption that either
$G_{T7}$ is zero (scenario I) or $G_{T5}$ (scenario II). All other transversity GPDs are neglected. Taking the
parameterization of the $p-p$ GPDs from previous work \ci{GK5,GK6} and evaluating the subprocess amplitudes
within the modified perturbative approach which effectively takes into account the transverse size of the
meson, we are in the position to predict the partial cross section for $\gamma^*p\to \pi^-\Delta^{++}$.

A precise calculation of observables for exclusive electroproduction of $\pi^-\Delta^{++}$ or for other
$\pi\Delta(1232)$ channels is beyond feasibility at present. There are many uncertainties: For instance
due to the use of large-$N_C$ results or due to the neglect of many of the $p-\Delta$ GPDs. Other uncertainties,
although of weaker importance, are the parameterizations of the $p-p$ GPDs or the exact treatment of the twist-3
contribution, e.g.\ the neglect of possible three-particle configurations of the meson state (see \ci{KP21}).
With regard to all these uncertainties we consider our investigation of exclusive electroproduction of
$\pi^-\Delta^{++}$ as a rough estimate. The trends and magnitudes are probably correct but not the details.
This is also the reason why we did not discuss asymmetries - with $A_{LL}$ as an exception.
Most of the asymmetries depend on the badly known relative phases of the helicity amplitudes.
With regard to the experimental program of the Jefferson lab our study seems to be timely.
Its results are perhaps useful as a starting point of a GPD analysis of data to come.

The extension of our study to other octet-decuplet transitions is straightforward with the help of SU(3)
flavor symmetry \ci{belitsky,frankfurt99}. Of course, this way also various octet-octet transitions GPDs are
related to the $p-p$ ones \ci{frankfurt}. Some of these relations have already been used to evaluate the
electroproduction cross sections for $K\Lambda$ and $ K\Sigma$ \ci{kroll19,GK6}.

%%%%%%%%%%%%%%%%%%%%%%%%%%%%%%%%%%%%%%%%%%%%%%%%%%%%%%%%%%%%%%%%%%%%%%%%%%%%%%%%%%%%%%%%%%%%%%%%%%%%%%

{\it Acknowledgment} 
We thank S.~Diehl for discussions and for informing us about the CLAS measurements of the $\pi^-\Delta^{++}$
asymmetry. This publication is supported by the Croatian Science Foundation project IP-2019-04-9709,
and by the EU Horizon 2020 research and innovation programme, STRONG-2020
project, under grant agreement No 824093.
   
%%%%%%%%%%%%%%%%%%%%%%%%%%%%%%%%%%%%%%%%%%%%%%%%%%%%%%%%%%%%%%%%%%%%%

\end{document}